\documentclass[a4paper]{article}

\usepackage[english]{babel}
\usepackage[utf8]{inputenc}
\usepackage[numbers,square,sort&compress]{natbib}

\usepackage{mathtools}
\usepackage{bm}
\usepackage{siunitx}
\usepackage{tabularx}
\usepackage{booktabs}
\usepackage[normalem]{ulem}
\usepackage{graphicx}
\usepackage[width=1.2\columnwidth,textfont=small]{caption}

\usepackage{amsmath,amssymb,multirow}

\usepackage{amsthm}

\DeclarePairedDelimiter{\normbracket}{\lVert}{\rVert}
\DeclarePairedDelimiter{\absbracket}{\lvert}{\rvert}

\DeclareMathOperator{\diag}{diag}

\newcommand{\abs}[1]{\absbracket*{#1}}
\newcommand{\normtwo}[1]{\normbracket*{#1}_2}
\newcommand{\transpose}{\ensuremath{\top}}

\newtheorem{defi}{Definition} 
\newtheorem{remark}[defi]{Remark}

\usepackage[colorlinks=true, allcolors=blue]{hyperref}

\title{\bf Set-point control and local stability for flat nonlinear systems using model-following control\thanks{Preprint provisionally accepted for publication by at-Automatisierungstechnik, De Gruyter}}

\author{
	Julian Willkomm\thanks{Control Engineering Group,  Technische Universt\"at Ilmenau, P.O.~Box 10 05 65, D-98684 Ilmenau, Germany.} \and
	Kai Wulff$^\dagger$\thanks{Corresponding author: \texttt{kai.wulff@tu-ilmenau.de}} \and 
	Johann Reger$^\dagger$}

\begin{document}

\maketitle

\begin{abstract}
We consider the set-point control problem for nonlinear systems with flat output that are subject to perturbations.
The nonlinear dynamics as well as the perturbations are locally Lipschitz.
We apply the model-following control (MFC) approach which consists of a model control loop (MCL) for a feedforward generation and a process control loop (PCL) that compensates the perturbations using high-gain feedback.
We analyse the resulting closed-loop system and discuss its relation to a standard flatness-based high-gain approach.
In particular we analyse the estimated region of attraction provided by a quadratic Lyapunov function.
A case study illustrates the approach and quantifies the region of attraction obtained for each control approach.
Using the initial condition of the model control loop as tuning parameter for the MFC design, provides that a significantly larger region of attraction can be guaranteed compared to a conventional single-loop high-gain design.
\end{abstract}

\section{Introduction}

\begin{figure*}[htb]
	\centering
	\includegraphics{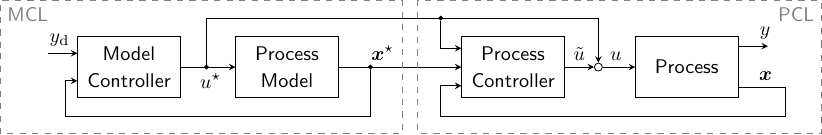}
	\caption{Block diagram model-following control (MFC)}
	\label{fig:blockschaltbild-mfc}
\end{figure*}

Model-following control (MFC), also referred to as model reference control, is a well-established control architecture consisting of a reference or model control loop (MCL) and the actual process control loop (PCL). A typical configuration of such structure is depicted in Figure~\ref{fig:blockschaltbild-mfc}.
This variation of the classical two-degree-of-freedom control structure \cite{Horowitz1963} has been applied and studied in many variations and control applications \cite{Erzberger1968,Ambrosino1985,Roppenecker1990,Wurmthaler2009,Osypiuk2010a,Huber2013,Brzozka2012}.

The MCL typically contains a linear model with a feedback controller 
that is often considered as imposing the desired behaviour on the control system.
The PCL 
is designed towards disturbance rejection.
However, since the reference control loop also provides a nominal control signal $u^\star$ and corresponding output $y^\star$ this control loop can also be considered as a feedforward control \cite{Wurmthaler2009} and trajectory generator \cite{Roppenecker2009,GraichenEK2010,SchaperASS2013}.

In many cases the MFC structure shows better robustness properties w.r.t. model uncertainties and disturbances compared to a single-loop feedback system, see e. g.
\cite{Skoczowski2003}.
Not least for this reason, a linear approximation of the nonlinear process is often used in the MCL in various control applications, e.\,g. a robotic manipulator \cite{Osypiuk2010,Osypiuk2010a,Osypiuk2010b}, the low-frequency motions of the drilling vessel \emph{Wimpey Sealab} \cite{Dworak2009}, an electric heater \cite{Skoczowski2001}, speed control of a permanent magnet synchronous motor  \cite{Pajchrowski2011} or boom cranes~\cite{SchaperDASS2014}.

Using a state-space model in the MCL the model-state $\bm{x}^\star$ is available as reference trajectory.
In \cite{Ambrosino1985,Roppenecker1990,Chern1998} tracking of the reference state generated by a linear feedback system is studied using state-feedback in the PCL.
The structure has been studied including a disturbance injection \cite{Roppenecker2009,Wurmthaler2009}.
In \cite{Brzozka2011} an MFC structure with a combination of output and state feedback is developed. 
Compared to the standard MFC structure the combined feedback shows better performance and robustness w.r.t parameter uncertainties. 
These results are obtained considering some case studies but a systematic analysis is not provided.

This paper contributes to the analysis of the MFC structure using a nonlinear model in the MCL. 
Such approach has been investigated in several case studies \cite{Pietrusewicz2012,Brzozka2012,Huber2013,Bauer2014,Schaper2014,Willkomm2019}
and shows good performance and robustness properties.
In \cite{Willkomm2018,Willkomm2019} a local model network to model the nonlinear process is used in the MCL. This feedforward control shows better performance than a comparable inversion-based approach.
Some initial analysis of the potential benefits using a nonlinear model in the MCL is found in \cite{Willkomm2021,Willkomm2022,Willkomm2023}.
In \cite{Willkomm2021} it is shown that the increase of robustness can be quantified in terms of the norm bound on the uncertainty.
In \cite{Willkomm2022} a high-gain control is applied to the PCL using feedback-linearisation based on the nominal nonlinear model in the MCL.
This approach shows significantly better robustness and performance properties, while requiring less control effort compared to a single-loop high-gain control design \cite{TietzeWR2024}.

In this contribution we consider nonlinear systems in Byrnes-Isidori form without internal dynamics but subject to locally Lipschitz uncertainties.
We investigate the capabilities of set-point control using the MFC scheme with feedback linearisation and high-gain state-feedback in the PCL.
In particular we study the robustness properties and estimates of the region of attraction for this control approach.
The benefits of the MFC structure with two control loops are highlighted and illustrated by a benchmark example.
Since the system exhibits a flat output we also discuss the relation of the MFC scheme to the well-known flatness-based 
approach \cite{Fliess1995}. 
Note however that the MFC scheme can also be applied to nonlinear systems with internal dynamics (i.e. the output is not flat), where the benefits of this structure are pronounced, see e.g. \cite{Willkomm2023,TietzeWR2024,TietzeWR2024b}.

The paper is structured as follows. The next section states the problem definition. 
In Section~\ref{sec:reglerentwurf-mfc} we present the proposed model-following control design and discuss its relation to a flatness-based approach. 
Section~\ref{sec:analysis} provides an analysis of the set-point control w.r.t. the steady-state error and stability. 
Furthermore, we discuss the results obtained by a single-loop design (flatness-based approach). 
In order to illustrate and quantify the effects on the estimated region of attraction we consider a standard mass-spring-damper system in Section~\ref{sec:case-study}.
Finally Section~\ref{sec:numerical-results} illustrates the results for several simulation scenarios.

\section{Problem definition}
\label{sec:problem-definition}

We consider a nonlinear SISO system in normal form with state vector $\bm x(t)\in\mathbb{R}^n$ and scalar input and output $u(t), y(t) \in\mathbb{R}$, respectively, given by 
\begin{align}\label{eq:sys_process}
	\dot{\bm{x}} &= \bm{A} \bm{x} + \bm{b} \big( f(\bm{x}) + g(\bm{x}) u + \phi(\bm{x}) \big) \\
	y &= \bm{c}^\transpose \bm{x},
\end{align}
with $\bm{A}\in\mathbb{R}^{n\times n}$, $\bm{b}\in\mathbb{R}^n$ in Brunovsk\'{y} form and $\bm{c}\in\mathbb{R}^n$ given by
\begin{align}\label{eq:def_Abc}
	\bm{A}&\!=\!\begin{pmatrix}
		0 & 1 & 0 & \cdots & 0 \\
		\vdots & \ddots & \ddots & \ddots & \vdots  \\
		\vdots & & \ddots & \ddots & 0 \\
		\vdots & &  & \ddots & 1 \\
		0 & \cdots& \cdots & \cdots  & 0  
	\end{pmatrix}\!\!,\! & 
	\begin{aligned}
		\bm{b} &\!=\! \begin{pmatrix}
			0 & \cdots & 0 &	1
		\end{pmatrix}^\transpose\!\!,  \\
		\bm{c}&\!=\!\begin{pmatrix}
			1 & 0 & \cdots &  0
		\end{pmatrix}^\transpose\!\!. 
	\end{aligned}
\end{align}
The structure implies that the output $y$ has relative degree~$n$ with respect to the input $u$ and hence $y$ is a flat output of the system.

The nonlinear functions $f,g: \mathbb{R}^n \to \mathbb{R}$ are known and all (deterministic) model uncertainties are represented by the matched uncertainty $\phi: \mathbb{R}^n \to \mathbb{R}$.
The functions $f,g,\phi$ shall be sufficiently smooth and locally Lipschitz for $\bm{x}\in\mathcal D\subset\mathbb{R}^n$ including the origin, and $g(\bm{x}) \neq 0$ for all $\bm{x}\in\mathcal D$.
The uncertainty $\phi$ shall satisfy  the Lipschitz condition 
\begin{equation}
	\normtwo{\phi(\bm{x}_a) - \phi(\bm{x}_b)} \leq \gamma \normtwo{\bm{x}_a - \bm{x}_b}  \label{eq:lipschitz-condition-uncertainty}
\end{equation}
for all $\bm{x}_a,\bm{x}_b\in\mathcal{D}$, where $\gamma> 0$ denotes the Lipschitz constant. 

The goal is to design a controller to asymptotically track a constant reference output $y_{\mathrm{d}}\in\mathbb{R}$ and thus the corresponding reference state $\bm{x}_{\mathrm{d}}= (y_{\mathrm{d}} \enspace 0 \enspace \cdots \enspace 0)^\transpose\in\mathbb{R}^n$.

\begin{remark}
	As this paper is focused on the treatment of flat systems, the process description in the Brunovsk\'y form \eqref{eq:sys_process} is canonical.
	Note however that the nonlinear MFC scheme discussed here is also applicable to systems for which no flat output exists \cite{Willkomm2022,Willkomm2023,TietzeWR2024,TietzeWR2024b}.
	In case the considered process is not in the form \eqref{eq:sys_process}, but exhibits a flat output $y_\mathrm{f}$, there exists a transformation into the form \eqref{eq:sys_process}.
	Note that for such case the form \eqref{eq:sys_process} exhibits matched uncertainties if and only if the uncertainties in the original coordinates are matched \cite{TietzeWR2025ECC}.
	Therefore we may restrict our analysis of flat systems to the system class~\eqref{eq:sys_process} without loss of generality.
\end{remark}

\section{Control design}
\label{sec:reglerentwurf-mfc}

While our main results focus on set-point control we will formally introduce the tracking control law for arbitrary $n$-times differentiable trajectories.
The design goal is to asymptotically track the desired output $y_{\mathrm{d}}(t)\in\mathbb{R}$, such that $\lim_{t\to\infty} y(t) - y_{\mathrm{d}}(t)=0$, while ensuring the boundedness of all reference $\bm{x}$. 
The desired output $y_{\mathrm{d}}(t)$ and its derivatives shall only be known during run-time such that inversion-based approaches, e.g. \cite{Graichen2005}, are not applicable.
The dynamics of the reference state $\bm{x}_{\mathrm{d}} = \big(y_{\mathrm{d}} \enspace \dot{y}_{\mathrm{d}} \enspace \dots y_{\mathrm{d}}^{(n-1)}\big) \in\mathbb{R}^n$ can be written as exo-system
\begin{equation}
	\dot{\bm{x}}_{\mathrm{d}} = \bm{A} \bm{x}_{\mathrm{d}} + \bm{b} y_{\mathrm{d}}^{(n)} \, , \qquad \bm{x}_{\mathrm{d}}(0) = \bm{x}_{\mathrm{d}, 0},
\end{equation}
where $y_{\mathrm{d}}^{(n)}$ denotes the $n$-th derivative of the reference trajectory and matrices $\bm{A}$ and $\bm{b}$ are given as in~\eqref{eq:def_Abc}.
For set-point control, i.e. a constant output reference $y_{\mathrm{d}}$, the reference state simplify to $\bm{x}_{\mathrm{d}} = (y_{\mathrm{d}} \enspace 0 \enspace \dots \enspace  0)^\transpose$ and satisfy $\bm{A} \bm{x}_{\mathrm{d}} = 0$.

We consider the 2DoF control architecture known as \emph{model-following control} (MFC) depicted in Figure~\ref{fig:blockschaltbild-mfc}. 
The model control loop (MCL) uses a nominal model of the process
\begin{equation*}
	\dot{\bm{x}}^\star = \bm{A} \bm{x}^\star + \bm{b} \big(f(\bm{x}^\star)+ g(\bm{x}^\star) u^\star\big) 
\end{equation*}
where $\bm{x}^\star(t) \in \mathbb{R}^n$ denotes the model state and \mbox{$u^\star(t) \in \mathbb{R}$} is the model control input. The goal of the model controller is to asymptotically stabilise the reference state $\bm{x}_{\mathrm{d}}$.
Furthermore the MCL provides $\bm{x}^\star$ as suitable reference signal to the PCL.

Defining the error state for the MCL $\tilde{\bm{x}}^\star \coloneqq \bm{x}^\star - \bm{x}_{\mathrm{d}}$ 
we obtain the error dynamics by
\begin{align*}
	\dot{\tilde{\bm{x}}}^\star &= \bm{A} \tilde{\bm{x}}^\star + \bm{b}\big(f(\bm{x}_{\mathrm{d}} + \tilde{\bm{x}}^\star) + g(\bm{x}_{\mathrm{d}} + \tilde{\bm{x}}^\star) u^\star -y_{\mathrm{d}}^{(n)} \big).
\end{align*}
We use the feedback linearising control law
\begin{align}
	u^\star = \frac{1}{g(\bm{x}_{\mathrm{d}} + \tilde{\bm{x}}^\star)} \left( -f(\bm{x}_{\mathrm{d}} + \tilde{\bm{x}}^\star) + y_{\mathrm{d}}^{(n)} + {\bm{k}^\star{}}^\transpose \tilde{\bm{x}}^\star  \right) \label{eq:model-controller}
\end{align}
with feedback gain ${\bm{k}^\star{}}^\transpose=(k_1^\star \enspace \dots \enspace k_n^\star) \in \mathbb{R}^{1\times n}$ and obtain the MCL closed loop dynamics 
\begin{align*}
	\dot{\tilde{\bm{x}}}^\star &= (\bm{A}+\bm{b}{\bm{k}^\star{}}^\transpose) \tilde{\bm{x}}^\star \, . 
\end{align*}
The feedback gain ${\bm{k}^\star{}}\in\mathbb{R}^{n}$ is designed such that $\bm{A}+\bm{b}{\bm{k}^\star{}}^\transpose$ is Hurwitz. 

For the process controller design we use the deviation of the model and process state. 
Defining the error state $\tilde{\bm{x}} \coloneqq \bm{x} - \bm{x}^\star = \bm{x} - \tilde{\bm{x}}^\star - \bm{x}_{\mathrm{d}}$ and considering the control signal $u=u^\star + \tilde{u}$, the error dynamics of the PCL are given by
\begin{multline*}
	\dot{\tilde{\bm{x}}} = \bm{A} \tilde{\bm{x}} + \bm{b} \bigl( \tilde{f}(\bm{x}_{\mathrm{d}}+\tilde{\bm{x}}^\star, \tilde{\bm{x}}, u^\star) \\ + g(\bm{x}_{\mathrm{d}}+\tilde{\bm{x}}^\star+\tilde{\bm{x}}) \tilde{u} + \phi(\bm{x}_{\mathrm{d}}+\tilde{\bm{x}}^\star+\tilde{\bm{x}}) \bigr)
\end{multline*}
where 
\begin{align*}
	\tilde{f}(\bm{x}_{\mathrm{d}}+\tilde{\bm{x}}^\star, \tilde{\bm{x}}, u^\star)&\coloneqq f(\bm{x}_{\mathrm{d}}+\tilde{\bm{x}}^\star+\tilde{\bm{x}}) - f(\bm{x}_{\mathrm{d}}+\tilde{\bm{x}}^\star) \\
	&\quad  + \left( g(\bm{x}_{\mathrm{d}}+\tilde{\bm{x}}^\star+\tilde{\bm{x}}) - g(\bm{x}_{\mathrm{d}}+\tilde{\bm{x}}^\star) \right) u^\star   \, .
\end{align*}
Again, we use a feedback linearizing control law
\begin{equation}
	\tilde{u} = \frac{1}{g(\bm{x}_{\mathrm{d}}+\tilde{\bm{x}}^\star+\tilde{\bm{x}})} \left( -\tilde{f}(\bm{x}_{\mathrm{d}}+\tilde{\bm{x}}^\star, \tilde{\bm{x}},u^\star) + \tilde{\bm{k}}^\transpose \tilde{\bm{x}} \right), \label{eq:process-controller}
\end{equation}
where $\tilde{\bm{k}}\in\mathbb{R}^{n}$ is designed such that $\bm{A}+\bm{b}\tilde{\bm{k}}^\transpose$ is Hurwitz. 

We note the dynamics and control law in terms of the error state $\tilde x^\star$ and $\tilde x$, respectively.
The overall control applied to the process is given by the composition $u=u^\star+\tilde u$.
Thus the MFC scheme provides two degrees of freedom such that the feedback gains $\bm{k}^\star,\tilde{\bm{k}}\in\mathbb{R}^{n}$ can be used to design the performance and robustness w.r.t. the model uncertainties $\phi$ separately.

\subsection{Relation to flatness based approach}
\label{sec:control-design-flatness-based}

In the following we shall relate the MFC to a typical 2DoF control design structure known as flatness-based control~\cite{Fliess1995}. 	
We discuss the relation of the MFC scheme to the feedback linearisation approach as well as to the exact feedforward linearisation \cite{Hagenmeyer2003,HagenmeyerZ2004}.

The feedback linearising control law in a single-loop design reads
	\begin{equation}\label{eq:u_flatnessbased}
		u = \frac{1}{g(\bm{x})} \left(-f(\bm{x}) + y_{\mathrm{d}}^{(n)} + \bm{k}^\transpose \left(\bm{x} - \bm{x}_{\mathrm{d}} \right)\right) \, .
	\end{equation}
This control law can be separated into two parts. The first part, $\frac{1}{g(\bm{x})} \big(\!-f(\bm{x}) \!+\! y_{\mathrm{d}}^{(n)}\big)$, is responsible for the tracking performance and the second part, $\frac{1}{g(\bm{x})} \bm{k}^\transpose (x-x_{\mathrm{d}})$, with feedback gain $\bm{k}\in\mathbb{R}^n$ is designed to compensate perturbations and initial deviations of the process to the reference state $\bm{x}_{\mathrm{d}}$.
	
The overall MFC control law $u$ composed by the model control law \eqref{eq:model-controller} and process control law \eqref{eq:process-controller} reads
\begin{align}
	u = u^\star + \tilde{u} &=  \frac{1}{g(\bm{x})} \Big( -f(\bm{x}) + y_{\mathrm{d}}^{(n)} \notag \\ 
	&\quad + {\bm{k}^\star{}}^\transpose \left(\bm{x}^\star - \bm{x}_{\mathrm{d}}\right) + \tilde{\bm{k}}^\transpose \left(\bm{x} - \bm{x}^\star \right)\Big) .
	\label{eq:MFC-controller-all}
\end{align} 
Compared with the feedback linearising control law \eqref{eq:u_flatnessbased} we see the similar structure but MFC exhibits two design parameters $\bm{k}^\star, \tilde{\bm{k}}\in\mathbb R^n$ whereas \eqref{eq:u_flatnessbased} has only one.
Typically, the feedback gain $\bm{k}^\star$ in \eqref{eq:model-controller} is used for the performance design including the transient behaviour of the initial deviation from the model state to the reference state $\bm{x}^\star(0)-\bm{x}_{\mathrm{d}}(0)$. The feedback gain $\tilde{\bm{k}}$ in \eqref{eq:process-controller} is designed to account for perturbations.

For the special case $\bm{x}^\star = \bm{x}$, the control law \eqref{eq:MFC-controller-all} reads
\begin{align}\label{eq:MFC-controller-all_reduced}
	u = \frac{1}{g(\bm{x})} \biggl( -f(\bm{x}) + y_{\mathrm{d}}^{(n)} + {\bm{k}^\star{}}^\transpose \left(\bm{x} - \bm{x}_{\mathrm{d}}\right)  \biggr),
\end{align}
which yields the control law of the classical flatness-based feedback linearising approach \eqref{eq:u_flatnessbased} with $\bm{k}=\bm{k}^\star$. 

The exact \emph{feedforward} linearisation control law \cite{Hagenmeyer2003} is given by
\begin{equation}\label{eq:u_ff_lin}
	u = \frac{1}{g(\bm{x}_{\mathrm{d}})} \left( -f(\bm{x}_{\mathrm{d}}) + y_{\mathrm{d}}^{(n)} + v_{\mathrm{fb}}
	\right),
\end{equation}
where $ v_{\mathrm{fb}}$ is some stabilising feedback.
Such control law can improve robustness with respect to model uncertainties upon  the feedback linearisation, \cite{Hagenmeyer2003}.
It shall be emphasised that 
the initial process state $\bm{x}(0)$ has to be sufficiently close to the reference state $\bm{x}_{\mathrm{d}}(0)$ in order to ensure stability \cite{Hagenmeyer2003}.
However in many cases the trajectory cannot be freely chosen, e.g. if it is provided by an operator or by a higher level hierarchical node.
In such scenarios the initial reference state may not be close enough to the process state. 
The MFC scheme relaxes this requirement, such that the initial process state $\bm{x}(0)$ may be in an arbitrary distance to the reference trajectory as long as the MCL can stabilise the reference trajectory.
In this sense the MCL generates a transition trajectory $\bm{x}^\star$ from its initial state to the reference trajectory $\bm{x}_\mathrm{d}$.

When applying the exact feedforward linearisation to the MCL we obtain the following special case.
Suppose the initial state of the reference trajectory and the initial model state are chosen identically $\bm{x}_{\mathrm{d}}(0)=\bm{x}^\star(0)$.
Then the model control law~\eqref{eq:model-controller} with $v_{\mathrm{fb}}=0$  reads
\begin{equation*}
u^\star = \frac{1}{g(\bm{x}^\star)} \left(-f(\bm{x}^\star) + y_{\mathrm{d}}^{(n)} \right),
\end{equation*}
which leads to $\bm{x}^\star(t)=\bm{x}_{\mathrm{d}}(t)$ for all $t\geq 0$.
Consequently, the MFC law~\eqref{eq:MFC-controller-all} simplifies to
\begin{equation*}
u = \frac{1}{g(\bm{x})} \left( -f(\bm{x}) + y_{\mathrm{d}}^{(n)} + \tilde{\bm{k}}^\transpose \left(\bm{x} - \bm{x}_{\mathrm{d}} \right) \right),
\end{equation*}
which matches the feedback linearising control law \eqref{eq:u_flatnessbased} with feedback $\bm{k}=\tilde{\bm{k}}$.
Thus in this case the MFC reduces to a flatness-based feedback linearising design.

\subsection{High-gain feedback design for the PCL}

Next we consider the feedback design of $\tilde{\bm{k}}$ for the process control loop (PCL).
A high-gain state feedback as the process controller is able to suppress the influence of the (matched) model uncertainty \cite{Khalil2002}. 
In the context of MFC this has been studied in \cite{Willkomm2022,TietzeWR2024}. 
Thus, we use a high-gain state feedback calculated by
\begin{equation*}
\tilde{\bm{k}}^\transpose={\bm{k}^\star{}}^\transpose \bm{D}^{-1} \varepsilon^{-1},
\end{equation*}
where $\bm{D}\coloneqq\diag(\varepsilon^{n-1}, \varepsilon^{n-2}, \dots, 1)$ and 
$0<\varepsilon \leq 1$. 
The parameter $\varepsilon$ can be interpreted as a time-scaling. 
The matrix $\bm{D}$ is a state-transformation producing the time-scaled state $\bm{z}\coloneqq \bm{D}^{-1} \tilde{\bm{x}}$.
Then the PCL reads
\begin{align*}
\dot{\bm{z}} \!=\! \big(\! \bm{D}^{-1}\! \bm{A} \bm{D} + \bm{D}^{-1} \bm{b} {\bm{k}^\star{}}^\transpose \!\varepsilon^{-1} \!\big) \bm{z}  + \bm{D}^{-1} \bm{b}\, \phi(\bm{x}_{\mathrm{d}} \!+\! \tilde{\bm{x}}^\star \!+\! \bm{D} \bm{z}).
\end{align*} 
Using \eqref{eq:def_Abc} we obtain $\bm{D}^{-1}\bm{b}=\bm{b}$ and $\varepsilon \bm{D}^{-1} \bm{A} \bm{D}=\bm{A}$, and therefore
\begin{align*}
\varepsilon \dot{\bm{z}} &= (\bm{A}+\bm{b}{\bm{k}^\star{}}^\transpose) \bm{z} + \varepsilon \bm{b}\, \phi(\bm{x}_{\mathrm{d}} + \tilde{\bm{x}}^\star + \bm{D} \bm{z}) \, .
\end{align*}
Combining both designs the closed-loop system dynamics of the MFC with the scaled state are given by
\begin{align}
\dot{\tilde{\bm{x}}}^\star &= (\bm{A}+\bm{b}{\bm{k}^\star{}}^\transpose) \tilde{\bm{x}}^\star \label{eq:closed-loop-sys-xtildestar} \\
\varepsilon \dot{\bm{z}} &= (\bm{A}+\bm{b}{\bm{k}^\star{}}^\transpose) \bm{z} + \varepsilon \bm{b}\, \phi(\bm{x}_{\mathrm{d}} + \tilde{\bm{x}}^\star + \bm{D} \bm{z}) \, . \label{eq:closed-loop-sys-z}
\end{align}

\section{Set-point control}
\label{sec:analysis}

In the following we consider the set-point control problem, i.e. $y_{\mathrm{d}}^{(i)}=0$ for $i=1,\ldots,n$, and $\bm{x}_{\mathrm{d}}=(y_{\mathrm{d}} \enspace 0 \enspace \dots \enspace 0)^\transpose$. We compare the steady-state error and the estimated region of attraction for the MFC scheme to the single-loop design. 

We show that the MFC scheme with a large enough high-gain feedback in the process control loop can globally stabilize the equilibrium near the reference state $\bm{x}_{\mathrm{d}}$ if the Lipschitz condition~\eqref{eq:lipschitz-condition-uncertainty} holds for all $\bm{x} \in \mathbb{R}^n$.
However, if condition~\eqref{eq:lipschitz-condition-uncertainty} is satisfied only locally, stability can only be ensured locally. 
For this case we propose an approach to compute the estimated region of attraction for the MFC scheme. 
Furthermore, we compare the region of attraction obtained for the MFC scheme to a corresponding estimate for a single-loop design, using a standard linear feedback as well as a high-gain feedback design.

\subsection{Steady-state error and stability analysis}
\label{sec:MFC_steady-state_and_stability}
We calculate the steady state in terms of the scaled state~$\bm{z}_{\mathrm{s}}$ by
\begin{align*}
	0 &= (\bm{A}+\bm{b}{\bm{k}^\star{}}^\transpose) \bm{z}_{\mathrm{s}} + \varepsilon \,\bm{b}\, \phi(\bm{x}_{\mathrm{d}}+\bm{D}\bm{z}_{\mathrm{s}}) 
\end{align*}
and obtain $z_{\mathrm{s}, i}=0$ for $i=2, \dots, n$. 
The first component $z_{\mathrm{s}, 1}$ is determined by
\begin{equation}
	0 = k_1^\star z_{\mathrm{s}, 1} + \varepsilon \phi(\bm{x}_{\mathrm{d}} + \bm{D} \bm{z}_{\mathrm{s}}) \, . \label{eq:steady-state-error-allg} 
\end{equation}
Thus, the steady state $\bm{z}_{\mathrm{s}}$ depends on the uncertainty $\phi(\bm{x})$, if
$\phi(\bm{x})$ depends on $x_1$ (and so $\phi(\bm{x}_{\mathrm{d}} + \bm{D} \bm{z}_{\mathrm{s}})$ dependents on $y_{\mathrm{d}} + z_{\mathrm{s}, 1}$). Otherwise the steady-state error is $\bm{z}_{\mathrm{s}}=0$.
Note that the uncertainty may cause multiple solutions of Eq.~\eqref{eq:steady-state-error-allg}. In such a case we refer to the solution closest to zero as $z_{\mathrm{s}, 1}$.
This effect is illustrated by the second simulation scenario in Section~\ref{sec:numerical-results}.

The unscaled error steady-state $\tilde{\bm{x}}_{\mathrm{s}}$ is equivalent to the stationary deviation from the process state to the reference state $\bm{x}_{\mathrm{s}}-\bm{x}_{\mathrm{d}}$, because $\bm{x}_{\mathrm{s}}=\bm{x}_{\mathrm{d}}+\tilde{\bm{x}}^\star_{\mathrm{s}}+\bm{D} \bm{z}_{\mathrm{s}}=\bm{x}_{\mathrm{d}}+\tilde{\bm{x}}_{\mathrm{s}}$.

For convenience of the stability analysis we shift the origin to the steady-state by introducing $\tilde{\bm{z}} \coloneqq \bm{z}-\bm{z}_{\mathrm{s}}$.
The dynamics in these coordinates are given by
\begin{align}
	\varepsilon \dot{\tilde{\bm{z}}} &= (\bm{A}+\bm{b} {\bm{k}^\star{}}^\transpose)\tilde{\bm{z}} + \varepsilon \bm{b}\, \tilde{\phi}(\bm{x}_{\mathrm{d}}+\bm{D} \bm{z}_{\mathrm{s}}, \tilde{\bm{x}}^\star + \bm{D} \tilde{\bm{z}} ), \label{eq:closed-loop-sys-ztilde}
\end{align} 
where
\begin{align*}
	\tilde{\phi}(\bm{x}_{\mathrm{d}}+\bm{D} \bm{z}_{\mathrm{s}}, \tilde{\bm{x}}^\star + \bm{D} \tilde{\bm{z}} ) \coloneqq 
	\phi(\bm{x}_{\mathrm{d}} + \tilde{\bm{x}}^\star + \bm{D} ( \bm{z}_{\mathrm{s}} + \tilde{\bm{z}} ) ) - \phi(\bm{x}_{\mathrm{d}} + \bm{D} \bm{z}_{\mathrm{s}})	 \, .
\end{align*}

Assuming the state remains within the domain for which~$\phi$ is Lipschitz, i.e. $\bm{x}=\bm{x}_{\mathrm{d}}+\tilde{\bm{x}}^\star+\bm{D} (\bm{z}_{\mathrm{s}}+\tilde{\bm{z}})\in\mathcal{D}$ and $\bm{x}_{\mathrm{s}}=\bm{x}_{\mathrm{d}} + \bm{D} \bm{z}_{\mathrm{s}}\in\mathcal{D}$
and with $\normtwo{\bm{D}}=1$ for $0<\varepsilon<1$ the following estimate holds:
\begin{align}
\normtwo{\tilde{\phi}(\bm{x}_{\mathrm{d}}+\bm{D} \bm{z}_{\mathrm{s}}, \tilde{\bm{x}}^\star + \bm{D} \tilde{\bm{z}})} &= \normtwo{\phi(\bm{x}_{\mathrm{d}} + \tilde{\bm{x}}^\star + \bm{D} ( \bm{z}_{\mathrm{s}} + \tilde{\bm{z}})) - \phi(\bm{x}_{\mathrm{d}} + \bm{D} \bm{z}_{\mathrm{s}}) } \notag \\
	&\leq \gamma \normtwo{\bm{x}_{\mathrm{d}} + \tilde{\bm{x}}^\star + \bm{D} ( \bm{z}_{\mathrm{s}} + \tilde{\bm{z}} )  - \bm{x}_{\mathrm{d}} - \bm{D} \bm{z}_{\mathrm{s}}} \notag \\
	&\leq  \gamma \normtwo{\tilde{\bm{x}}^\star + \bm{D} \tilde{\bm{z}}} \notag \\
	&\leq \gamma \left( \normtwo{\tilde{\bm{x}}^\star} + \normtwo{\tilde{\bm{z}}} \right)\,. \label{eq:Lipschitz-phitilde}
\end{align}

Consider the block-diagonal quadratic Lyapunov function
\begin{equation}
	V(\tilde{\bm{x}}^\star,\tilde{\bm{z}})=\vartheta {\tilde{\bm{x}}^\star{}}^\transpose \bm{P} \tilde{\bm{x}}^\star + \tilde{\bm{z}}^\transpose \bm{P} \tilde{\bm{z}}, \label{eq:LF-MFC}
\end{equation}
where the scalar $\vartheta>0$ is a weight of the error state in the MCL and will be determined later. 
The positive definite, symmetric matrix $\bm{P}\in\mathbb{R}^{n\times n}$ satisfies the Lyapunov equation
\begin{equation}\label{eq:LF_equation}
	\big(\bm{A}+\bm{b}{\bm{k}^\star{}}^\transpose\big)^\transpose \bm{P} + \bm{P}\big(\bm{A}+\bm{b}{\bm{k}^\star{}}^\transpose\big) = -\bm{I},
\end{equation}
where $\bm{I}$  denotes the identity matrix of appropriate dimension.
Note that the same matrix $\bm{P}$ is used for both quadratic parts in the Lyapunov function~\eqref{eq:LF-MFC}.
		
The derivative of the Lyapunov function along the solution of \eqref{eq:closed-loop-sys-xtildestar} and \eqref{eq:closed-loop-sys-ztilde} yields
\begin{align*}
			\dot{V}(\tilde{\bm{x}}^\star, \tilde{\bm{z}}) =- \vartheta {\tilde{\bm{x}}^\star{}}^\transpose \tilde{\bm{x}}^\star - \varepsilon^{-1} \tilde{\bm{z}}^\transpose \tilde{\bm{z}}  + 2 \tilde{\phi}(\bm{x}_{\mathrm{d}}+\bm{D} \bm{z}_{\mathrm{s}}, \tilde{\bm{x}}^\star + \bm{D} \tilde{\bm{z}}) \bm{b}^\transpose \bm{P} \tilde{\bm{z}}
\end{align*}
and can be bounded by
		\begin{align*}
			\dot{V}(\tilde{\bm{x}}^\star, \tilde{\bm{z}}) \leq -\vartheta \normtwo{\tilde{\bm{x}}^\star}^2 - \varepsilon^{-1} \normtwo{\tilde{\bm{z}}}^2  + 2 \normtwo{\tilde{\phi}(\bm{x}_{\mathrm{d}}+\bm{D} \bm{z}_{\mathrm{s}}, \tilde{\bm{x}}^\star + \bm{D} \tilde{\bm{z}})} \normtwo{\bm{b}^\transpose \bm{P}} \normtwo{\tilde{\bm{z}}} \, .
		\end{align*}
		Substituting the estimate \eqref{eq:Lipschitz-phitilde} yields
		\begin{align*}
			\dot{V}(\tilde{\bm{x}}^\star, \tilde{\bm{z}}) &\leq -\vartheta \normtwo{\tilde{\bm{x}}^\star}^2 - (\varepsilon^{-1} - 2 \gamma \normtwo{\bm{b}^\transpose \bm{P}} ) \normtwo{\tilde{\bm{z}}}^2   + 2 \gamma \normtwo{\bm{b}^\transpose \bm{P}} \normtwo{\tilde{\bm{x}}^\star} \normtwo{\tilde{\bm{z}}} \\
			&\leq - \mathcal{\bm{Y}}^\transpose \bm{M} \mathcal{\bm{Y}},
		\end{align*}
		where $\mathcal{\bm{Y}}:=\big(\normtwo{\tilde{\bm{x}}^\star}, \enspace \normtwo{\tilde{\bm{z}}}\big)^\transpose$ and 
		\begin{equation*}
			\bm{M}:= \begin{pmatrix}
				\vartheta & -\gamma \normtwo{\bm{b}^\transpose \bm{P}} \\
				-\gamma \normtwo{\bm{b}^\transpose \bm{P}} & \varepsilon^{-1} - 2 \gamma \normtwo{\bm{b}^\transpose \bm{P}}
			\end{pmatrix}.
		\end{equation*}
		For asymptotic stability, we require the matrix $\bm{M}$ to be strictly positive definite, i.e. all leading principal minors of $\bm{M}$ have to be positive.
		In particular we require $$\vartheta \varepsilon^{-1} - 2 \vartheta  \gamma \normtwo{\bm{b}^\transpose \bm{P}} - \gamma^2 \normtwo{\bm{b}^\transpose \bm{P}}^2>0,$$ and solving this for the uncertainty bound~$\gamma$ yields
		\begin{equation}
			\gamma < \frac{1}{\varepsilon \left( 1 + \sqrt{1 + (\vartheta \varepsilon)^{-1}} \right) \normtwo{\bm{b}^\transpose \bm{P}} }\eqqcolon\Gamma_{\mathrm{MFC}}\,. \label{eq:stabi-bedingung}
		\end{equation}
		We denote this upper bound for $\gamma$ as robustness bound~$\Gamma_{\mathrm{MFC}}$, which guarantees stabilisation of this class of uncertainties~$\phi$.
		The robustness bound depends on the time-scaling $\varepsilon$ and the weight $\vartheta$ of the Lyapunov function.
		For any given uncertainty bound $\gamma$ we can choose~$\varepsilon$ small enough such that the stability condition~\eqref{eq:stabi-bedingung} is satisfied.
		For $\vartheta\rightarrow\infty$ we obtain 
		\begin{align}\label{eq:Gamma-sl-comparison-allg}
			\Gamma_{\mathrm{MFC}} &=  \frac{1}{2 \varepsilon \normtwo{\bm{b}^\transpose \bm{P}}} \, .
		\end{align}

		Of course the robustness analysis can either be conducted in order to enlarge the estimated region of attraction for a fixed $\gamma$ (see Section \ref{sec:case-study}), or to enlarge the maximum robustness bound $\Gamma_{\mathrm{MFC}}$ for a fixed estimated region of attraction.

		In order to facilitate comprehension of the manuscript we summarise some of the variables used for reference in Table \ref{tab:states}.
		With these relations the considered equilibrium is given by
		$\begin{pmatrix}
			{\bm{x}^\star}^\top &
			\bm{x}^\top
		\end{pmatrix}^\top = \begin{pmatrix}
			\bm{x}_{\mathrm{d}}^\top
			&
			\bm{x}_{\mathrm{s}}^\top
		\end{pmatrix}^\top,
		$
		whereas the uncertainty $\tilde{\phi}$ in Eq.~\eqref{eq:closed-loop-sys-ztilde} in original coordinates reads 
		\begin{align*}
			\tilde{\phi}(\bm{x}_{\mathrm{s}}, \bm{x}-\bm{x}_\mathrm{s}) &= \phi(\bm{x}) - \phi(\bm{x}_{\mathrm{s}}) \,. 
		\end{align*}
		\begin{table}[htb]
			\centering
			\begin{tabularx}{0.65\linewidth}{Xl}
				\toprule
				{\small Process state} & $\bm{x}=\bm{x}_{\mathrm{d}} + \tilde{\bm{x}}^\star + \bm{D}\left(\bm{z}_{\mathrm{s}} + \tilde{\bm{z}}\right)$ \\
				{\small Model error state} & $\tilde{\bm{x}}^\star\!\coloneqq\bm{x}^\star-\bm{x}_{\mathrm{d}}$ \\
				{\small Steady-state} & $\bm{x}_{\mathrm{s}}\!=\bm{x}_{\mathrm{d}} + \bm{D} \bm{z}_{\mathrm{s}}$ \\
				{\small Process error state} & $\tilde{\bm{x}}\coloneqq \bm{x}-\bm{x}^\star=\bm{D}\bm{z}=\bm{D}\left(\bm{z}_{\mathrm{s}} + \tilde{\bm{z}}\right)$ \\ 
				{\small Time-scaled state} & $\bm{z}\coloneqq \bm{D}^{-1}\tilde{\bm{x}}$ \\ 
				{\small Shifted scaled error state} & 
				$\tilde{\bm{z}}\coloneqq \bm{z}-\bm{z}_\mathrm{s}$\\
				&$\tilde{\bm{z}}=\bm{D}^{-1} \left( \left( \bm{x} - \bm{x}_{\mathrm{s}} \right) - \left( \bm{x}^\star - \bm{x}_{\mathrm{d}} \right) \right)$ \\ 
				\bottomrule
			\end{tabularx}
			\caption{Definitions and relations of some important states.}
			\label{tab:states}
		\end{table}

		\subsection{Comparison to single-loop designs}
		\label{sec:analyse-sl-slhg}
		For comparison, we consider system \eqref{eq:sys_process} with the single-loop control law
		\begin{align}
			u_{\mathrm{SL}} = \frac{1}{g(\bm{x})} \left( -f(\bm{x}) + \bm{k}_{\mathrm{SL}}^\transpose (\bm{x} - \bm{x}_{\mathrm{d}}) \right) \label{eq:single-loop-control-law-allg}
		\end{align} 
		and feedback gain $\bm{k}_{\mathrm{SL}}^\transpose$ chosen $\bm{k}_{\mathrm{SL}}^\transpose = {\bm{k}^\star{}}^\transpose$. The closed-loop dynamics are
		\begin{equation*}
			\dot{\bm{x}} = \big(\bm{A}+\bm{b}\bm{k}_{\mathrm{SL}}^\transpose\big) (\bm{x} - \bm{x}_{\mathrm{d}}) + \bm{b} \phi(\bm{x}) \, .
		\end{equation*}
		The steady-state is given by
		\begin{equation}
			0 = \big(\bm{A}+\bm{b}\bm{k}_{\mathrm{SL}}^\transpose\big) (\bm{x}_{\mathrm{s}} - \bm{x}_{\mathrm{d}}) + \bm{b} \phi(\bm{x}_{\mathrm{s}}) \label{eq:single-loop-steady-state-allg}
		\end{equation}
		and yields $x_{\mathrm{s}, i}=0$ for $i=2, \dots, n$ and 
		\begin{equation*}
			0 = k_1^\star (x_{\mathrm{s}, 1} - y_{\mathrm{d}}) + \phi(\bm{x}_{\mathrm{s}}).
		\end{equation*}
		Similar as in \eqref{eq:steady-state-error-allg} there may be several solutions depending on the uncertainty $\phi$. In the following we consider $\bm{x}_{\mathrm{s}}$ as the solution closest to $\bm{x}_\mathrm{d}$. 
		We define the error state $\bm{x}_e \coloneqq \bm{x}- \bm{x}_{\mathrm{s}}$ and calculate the error dynamics given by
		\begin{equation}
			\dot{\bm{x}}_e = (\bm{A}+\bm{b}\bm{k}_{\mathrm{SL}}^\transpose) \bm{x}_e + \bm{b} \tilde{\phi}(\bm{x}_ {\mathrm{s}}, \bm{x}_e), \label{eq:single-loop-error-dynamics}
		\end{equation} 
		where $\tilde{\phi}(\bm{x}_ {\mathrm{s}}, \bm{x}_e) \coloneqq \phi(\bm{x}_e + \bm{x}_{\mathrm{s}}) - \phi(\bm{x}_{\mathrm{s}})$.
		
		Assuming that $\bm{x}=\bm{x}_{\mathrm{s}}+\bm{x}_e \in\mathcal{D}$ and $\bm{x}_{\mathrm{s}}\in\mathcal{D}$ 
		we follow the stability analysis of the MFC scheme. 
		We choose a quadratic Lyapunov function 
		\begin{equation}
			V_{\mathrm{SL}}(\bm{x}_e) = \bm{x}_e^\transpose \bm{P} \bm{x}_e \label{eq:single-loop-Lyapunov-function}
		\end{equation} 
		with positive definite, symmetric matrix $\bm{P}$ as solution of the Lyapunov equation \eqref{eq:LF_equation}. 
		Since  $\bm{k}_{\mathrm{SL}}^\transpose = {\bm{k}^\star{}}^\transpose$ we obtain the same matrix $\bm{P}$ as before.
		The derivative of the Lyapunov function yields
		\begin{equation*}
			\dot{V}_{\mathrm{SL}}(\bm{x}_e) = -\bm{x}_e^\transpose \bm{x}_e + 2 \tilde{\phi}(\bm{x}_ {\mathrm{s}}, \bm{x}_e) \bm{b}^\transpose \bm{P} \bm{x}_e \, .
		\end{equation*}
		Using the Lipschitz condition~\eqref{eq:lipschitz-condition-uncertainty} of the model uncertainty 
		\begin{align*}
			\normtwo{\tilde{\phi}(\bm{x}_ {\mathrm{s}}, \bm{x}_e)} &\leq \gamma \normtwo{\bm{x}_e}
		\end{align*}
		and using further simplifications leads to 
		\begin{align*}
			\dot{V}_{\mathrm{SL}}(\bm{x}_e) &\leq - \left(1-2\gamma \normtwo{\bm{b}^\transpose \bm{P}}\right) \normtwo{\bm{x}_e}^2 \, .
		\end{align*}
		For asymptotic stability we require 
		\begin{equation}\label{eq:Gamma_condition_SL}
			\gamma<\Gamma_{\mathrm{SL}} \coloneqq \frac{1}{2 \normtwo{\bm{b}^\transpose \bm{P}}} \, .
		\end{equation}
		A comparison of the robustness bound of the single-loop design $\Gamma_{\mathrm{SL}}$ to the robustness bound of the MFC scheme $\Gamma_{\mathrm{MFC}}$ with a large parameter $\vartheta$ in \eqref{eq:Gamma-sl-comparison-allg} shows that $\Gamma_{\mathrm{MFC}}$ is approximately $\varepsilon^{-1}$ times as large as $\Gamma_{\mathrm{SL}}$ and thus allows for considerably larger uncertainties (or for a larger region of attraction for a fixed bound $\gamma$).

	Secondly, we consider a high-gain state-feedback in the single loop and choose $\bm{k}_{\mathrm{SL}}=\tilde{\bm{k}}$.
	Following the analysis above yields the steady-state in \eqref{eq:single-loop-steady-state-allg} with $x_{\mathrm{s}, i}=0$ for $i=2,\dots,n$ and 
	\begin{equation*}
		0=\tilde{k}_1 (x_{\mathrm{s}, 1} - y_{\mathrm{d}}) + \phi(\bm{x}_{\mathrm{s}}) \, .
	\end{equation*}
	Note that the high-gain state feedback $\bm{k}_{\mathrm{SL}}=\tilde{\bm{k}}$ in single loop leads to exactly the same steady-state $\bm{x}_{\mathrm{s}}$ as for the MFC scheme obtained from \eqref{eq:steady-state-error-allg} as some basic calculations reveal. 
	
	For the closed-loop analysis we define the scaled error state $\bm{z}_{\mathrm{SL}}\coloneqq \bm{D}^{-1} \left(\bm{x}-\bm{x}_{\mathrm{s}}\right)$ with the scaling matrix $\bm{D}=\diag\left(\varepsilon^{n-1}, \varepsilon^{n-2}, \dots, 1\right)$ and $0<\varepsilon\leq 1$. 
	Note that the scaling parameter $\varepsilon$ and matrix $\bm{D}$ are exactly the same as for the process control loop in the MFC scheme.
	But $\bm{z}_{\mathrm{SL}}$ is defined by the deviation of the process state from the steady-state.
	The closed-loop dynamics are given by 
	\begin{align}
		\varepsilon \dot{\bm{z}}_{\mathrm{SL}} &= \left(\bm{A}+\bm{b}{\bm{k}^\star}^\transpose\right) \bm{z}_{\mathrm{SL}} + \varepsilon \bm{b} \tilde{\phi}(\bm{x}_{\mathrm{s}}, \bm{D} \bm{z}_{\mathrm{SL}}). \label{eq:single-loop-high-gain-error-dynamics}
	\end{align}
	The Lyapunov function is chosen as 
	\begin{equation*}
		V_{\mathrm{SLHG}}(\bm{z}_{\mathrm{SL}}) = \bm{z}_{\mathrm{SL}}^\transpose \bm{P} \bm{z}_{\mathrm{SL}},
	\end{equation*}
	where the matrix $\bm{P}$ is given by~\eqref{eq:LF_equation} and is exactly the same solution as for the MFC scheme. 
	Using the Lipschitz condition~\eqref{eq:lipschitz-condition-uncertainty} with some basic calculations
	\begin{align*}
		\normtwo{\tilde{\phi}(\bm{x}_{\mathrm{s}}, \bm{D}\bm{z}_{\mathrm{SL}})} \leq \gamma \normtwo{\bm{D}\bm{z}_{\mathrm{SL}}} \leq \gamma \normtwo{\bm{z}_{\mathrm{SL}}}
	\end{align*}
	we obtain for an estimate of the derivative of the Lyapunov function
	\begin{equation*}
		\dot{V}_{\mathrm{SLHG}}(\bm{z}_{\mathrm{SL}}) \leq -\left(\varepsilon^{-1}-2\gamma\normtwo{\bm{b}^\transpose \bm{P}}\right) \normtwo{\bm{z}_{\mathrm{SL}}}^2.
	\end{equation*}
	Consequently, the robustness bound is given by 
	\begin{equation}
		\Gamma_{\mathrm{SLHG}} = \frac{1}{2 \varepsilon \normtwo{\bm{b}^\transpose \bm{P}}} 
		\label{eq:Gamma-SLHG-scaled}
	\end{equation}
	for the single-loop high-gain design. 
	
	A comparison shows that the robustness bounds of the single-loop high-gain design \eqref{eq:Gamma-SLHG-scaled} is the limit of the robustness bound of the MFC scheme \eqref{eq:Gamma-sl-comparison-allg} for $\vartheta\rightarrow\infty$.
	
	Typically the high-gain approach leads to a small steady-state error and a larger robustness bound at the expense of large initial control effort known as peaking phenomenon.
	It has been shown in \cite{TietzeWR2024} that this peaking-phenomenon can be completely avoided by an appropriate choice of the initial condition $\bm x^\star_0$ in den MCL when using high-gain feedback in the MFC scheme.

\begin{remark}
		Note that other choices for the feedback gain $\bm{k}_{\mathrm{SL}}$ in the single-loop design are applicable and result in the same steady-state error as the for the suggested MFC scheme. 
		In fact, basic calculations using Eqs.~\eqref{eq:steady-state-error-allg} and \eqref{eq:single-loop-steady-state-allg} reveal that any stabilising feedback gain $\bm{k}_{\mathrm{SL}}$ with $\tilde{k}_1$ as first element leads to the same steady-state error. However, applying another feedback gain affects the dynamics, the range of the control signal, and the robustness bound of the system and ultimately renders a comparison of the two control schemes more difficult.
\end{remark}

\begin{remark}
		This contribution considers static state-feedback only.
		Of course integral action in the control loop is capable of eliminating the steady-state error completely.
		While adding an integrator is a natural design approach, the augmented system dynamics have to be considered in the estimate of the region of attraction and will result in more complex and higher-order analysis.
		This is beyond the scope of this contribution, but shall be subject of future research.
\end{remark}

While the robustness bounds obtained for the single-loop high-gain and the MFC scheme are closely related, the analysis for their region of attraction is more involved since the Lyapunov functions $V_{\mathrm{SLHG}}$ and $V$ in \eqref{eq:LF-MFC} are defined in different coordinates.
Therefore we shall use the case study in Section \ref{sec:case-study} to illustrate the robustness properties and resulting estimates of the regions of attraction for the two approaches.

\section{Case study}
\label{sec:case-study}

\begin{figure}[htb]
	\centering
	\includegraphics{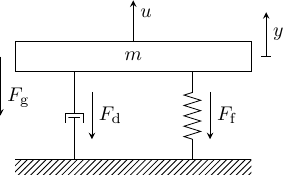}
	\caption{Sketch of the mass-springer-damper system.}
	\label{fig:feder-masse-system-skizze}
\end{figure}

We consider a standard mass-spring-damper process as depicted in Figure~\ref{fig:feder-masse-system-skizze}.
It consists of a mass $m$ supported by a hardening spring and a linear damper.
The system can be actuated by the force $u$ in order to regulate the vertical position $y$.

We obtain the process dynamics considering the balance of forces
\begin{equation}\label{eq:balance_of_forces_mass_spring_damper}
	m \ddot{y} + F_{\mathrm{f}} + F_{\mathrm{d}} + F_{\mathrm{g}} = u \, .
\end{equation}
The hardening spring generates the spring force
\begin{equation*}
	F_{\mathrm{f}}=k \big(1 + \alpha^2 y^2\big) y,
\end{equation*}
where $\alpha,k>0$ are the hardening factor and linear spring coefficient, respectively.
The damping shall be linear
\begin{equation*}
	F_{\mathrm{d}}=c_{\mathrm{d}} \dot{y}
\end{equation*}
with damping coefficient $c_{\mathrm{d}}>0$,
the gravitational force is
\begin{equation*}
	F_{\mathrm{g}} = m g_0,
\end{equation*}
where $g_0$ denotes the gravity constant.
Substitution of the forces into \eqref{eq:balance_of_forces_mass_spring_damper} yields the process dynamics
\begin{equation}\label{eq:DLG_mass_spring_damper}
	m \ddot{y} + c_{\mathrm{d}} \dot{y}+ k \big(1 + \alpha^2 y^2\big) y  + m g_0 = u \, .
\end{equation}
We consider the state vector $\bm{x}(t)=(y(t) \enspace \dot{y}(t))^\transpose \in \mathbb{R}^2$, the actuation force as input $u(t)\in\mathbb{R}$ and the displacement as output $y(t)\in\mathbb{R}$. 
The state space model is given by
\begin{align}
	\dot{\bm{x}} &= \bm{A} \bm{x} + \bm{b} (f(\bm{x})+ g(\bm{x}) u) \, , \\
	y&=\bm{c}^\transpose \bm{x}
\end{align}
with 
\begin{align*}
	\bm{A}&= \begin{pmatrix}
		0 & 1 \\
		0 & 0
	\end{pmatrix},\;\;  \bm{b}= \begin{pmatrix}
		0 \\
		1
	\end{pmatrix},\;\; 
	\bm{c}^\transpose = \begin{pmatrix}
		1 & 0
	\end{pmatrix},\;\;  g(\bm{x})= \frac{1}{m}, \\[1ex]
	f(\bm{x}) &=\mathmakebox[2em][l]{-\tfrac{k}{m}\big(1+\alpha^2 x_1^2\big)x_1 -\tfrac{c_{\mathrm{d}}}{m}\, x_2 - g_0.}
\end{align*}

The nominal parameters $\alpha$, $k$, $c_\mathrm{d}$ of a real process are usually not exactly known. Therefore, we consider parameter uncertainties for the parameters of the spring constant $\Delta k$, the hardening factor $\Delta \alpha$, and the damping coefficient $\Delta c_{\mathrm{d}}$.

Then the considered process dynamics are given by
\begin{subequations}\label{eq:msd-dgl-x}
	\begin{align}
	\dot{\bm{x}} &= \bm{A} \bm{x} + \bm{b} (f(\bm{x}) + g(\bm{x}) u + \phi(\bm{x})) \, , \\ y&=\bm{c}^\transpose \bm{x},
\end{align}
\end{subequations}
with locally Lipschitz uncertainty 
\begin{align}\label{eq:msd-uncertainty}
	\phi(\bm{x}) = - \frac{\Delta k}{m} \big( \alpha + \Delta \alpha\big)^2 x_1^3 - \frac{k}{m} \Delta \alpha \big( 2 \alpha + \Delta \alpha \big) x_1^3 - \frac{\Delta k}{m} x_1 - \frac{\Delta c_{\mathrm{d}}}{m} x_2.
\end{align}

\subsection{MFC design and stability analysis}

The controller for MFC scheme \eqref{eq:model-controller}--\eqref{eq:process-controller} is designed such that the eigenvalues of the MCL dynamics are all placed at $-2$ and the tuning parameter $\varepsilon$ is chosen as $\varepsilon=\num[round-mode=none]{0.1}$. The state feedback gains result in
\begin{subequations}\label{eq:control_parameter}
	\begin{align}
		{\bm{k}^\star{}}^\transpose&=\begin{pmatrix}
			-4 & -4
		\end{pmatrix},\\
		\tilde{\bm{k}}^\transpose &= {\bm{k}^\star{}}^\transpose \bm{D}^{-1} \varepsilon^{-1} = \begin{pmatrix}
			-400 & -40
		\end{pmatrix},  
	\end{align}
\end{subequations}
where $\bm{D}=\diag(\varepsilon, 1)$.
The system dynamics of the MFC system with scaled states are given as in \eqref{eq:closed-loop-sys-xtildestar}--\eqref{eq:closed-loop-sys-z}. 

Next we calculate the steady state of the scaled states $\bm{z}_{\mathrm{s}}$ and bring the dynamics into the desired closed-loop MFC form \eqref{eq:closed-loop-sys-xtildestar} and \eqref{eq:closed-loop-sys-ztilde}. 
The nominal system \eqref{eq:msd-dgl-x} without uncertainty and feedback \eqref{eq:control_parameter} has a unique equilibrium at the set-point $y_\mathrm{d}$.
Due to the third-order polynomial uncertainty $\phi(\bm{x})$, multiple equilibria with interchanging stability properties may occur (see Figure~\ref{fig:steady-state-solution} in Section \ref{sec:numerical-results} for an illustration).

From \eqref{eq:steady-state-error-allg}  with $z_{\mathrm{s}, 2}=0$  for the scaled steady-state~$\bm{z}_{\mathrm{s}}$, we obtain
\begin{equation*}
	\begin{multlined}
		0= k_1^\star z_{\mathrm{s}, 1} - \varepsilon \Big( \frac{\Delta k}{m} \big( \alpha + \Delta \alpha\big)^2 \big(y_{\mathrm{d}} + \varepsilon z_{\mathrm{s}, 1}\big)^3   + \frac{k}{m} \Delta \alpha \big( 2 \alpha + \Delta \alpha \big) \big(y_{\mathrm{d}} + \varepsilon z_{\mathrm{s}, 1}\big)^3 + \frac{\Delta k}{m} \big(y_{\mathrm{d}} + \varepsilon z_{\mathrm{s}, 1}\big) \Big),
	\end{multlined}
\end{equation*}
where $k_1^\star=-4$ is the first element of the feedback gain~$\bm{k}^\star$.
	Substituting $\varepsilon z_1=\tilde x_1$ yields the steady-state of the unscaled error $\tilde{\bm{x}}_{\mathrm{s}}=\bm{x}_{\mathrm{s}} - \bm{x}_{\mathrm{d}}$.
	The equilibria are given by the roots of the polynomial
	\begin{multline}
		p_{\mathrm{MFC}, \mathrm{s}}(\tilde{x}_1) := k_1^\star \varepsilon^{-2} \tilde{x}_{1} - \Big( \frac{\Delta k}{m} \big( \alpha + \Delta \alpha\big)^2 (y_{\mathrm{d}} + \tilde{x}_{1})^3  \\ 
		+ \frac{k}{m} \Delta \alpha \big( 2 \alpha + \Delta \alpha \big) \big(y_{\mathrm{d}} + \tilde{x}_{1}\big)^3 + \frac{\Delta k}{m} \big(y_{\mathrm{d}} + \tilde{x}_{1}\big) \Big). \label{eq:mfc-steady-state-function}
	\end{multline}
	Depending on the system parameters and the set-point some equilibria may vanish, see Figure \ref{fig:steady-state-solution} in Section \ref{sec:numerical-results}.

	Let us determine the uncertainty $\tilde{\phi}$. Therefore, we first consider the unscaled steady-state $\bm{x}_{\mathrm{s}}$ together with the unscaled error state $\bm{x}_e$ and calculate
	\begin{align}
		\tilde{\phi}(\bm{x}_{\mathrm{s}}, \bm{x}_e) &= 	
			- \frac{\Delta k}{m} x_{e, 1} - \frac{\Delta c_{\mathrm{d}}}{m} x_{e, 2}  
			- \frac{\sigma_1}{m} \left( \left(  x_{e, 1}  +  x_{\mathrm{s}, 1}  \right)^3  - x_{\mathrm{s}, 1}^3 \right)
	\notag \\[1ex]
		&= 	- \frac{\Delta k}{m} x_{e, 1} - \frac{\Delta c_{\mathrm{d}}}{m} x_{e, 2} - \frac{\sigma_1}{m} \left( 3 x_{\mathrm{s}, 1}^2  + 3 x_{\mathrm{s}, 1} x_{e, 1} + x_{e, 1}^2 \right) x_{e, 1}
		\notag \\[1ex]
		&= 	- \frac{\Delta k}{m} x_{e, 1} - \frac{\Delta c_{\mathrm{d}}}{m} x_{e, 2} 
			- \frac{\sigma_1}{m} \left( \left(  x_{e, 1}  + \frac{3}{2} x_{\mathrm{s}, 1}  \right)^2  + \frac{3}{4} x_{\mathrm{s}, 1}^2 \right) x_{e, 1} 
	\label{eq:msd-phi-tilde-xs-xe}
	\end{align}
	with the auxiliary variable 
	\begin{equation*}
		\sigma_1 = \Delta k\left(\alpha + \Delta \alpha\right)^2 + k \Delta \alpha \left(2\alpha + \Delta \alpha\right) \, .
	\end{equation*}
	Substituting $\bm{x}_{\mathrm{s}}=\bm{x}_{\mathrm{d}} + \bm{D} \bm{z}_{\mathrm{s}}$ and $\bm{x}_e=\tilde{\bm{x}}^\star+\bm{D}\tilde{\bm{z}}$, the uncertainty $\tilde{\phi}$ reads
	\begin{multline}
		\tilde{\phi}(\bm{x}_{\mathrm{d}}+\bm{D} \bm{z}_{\mathrm{s}}, \tilde{\bm{x}}^\star + \bm{D} \tilde{\bm{z}}) = 
		- \frac{\Delta k}{m} \left( \tilde{x}_1 + D_1 \tilde{z}_1  \right) - \frac{\Delta c_{\mathrm{d}}}{m} \left( \tilde{x}_2 + D_2 \tilde{z}_2  \right) \\
		- \frac{\sigma_1}{m} \Big( \big( \left(\tilde{x}_{1}^\star + D_1 \tilde{z}_{1} \right) + \tfrac{3}{2} \left(x_{\mathrm{d}, 1} + D_1 z_{\mathrm{s}, 1} \right)  \big)^2 
		+ \tfrac{3}{4} \big(x_{\mathrm{d}, 1} + D_1 z_{\mathrm{s}, 1} \big)^2 \Big) \left(\tilde{x}_1^\star + D_1 \tilde{z}_1 \right), \label{eq:msd-phi-tilde-xtildestar-ztilde}
	\end{multline}
	where $D_i$ with $i\in\{1,2\}$ denotes the $i$-th element on the diagonal of $\bm{D}$ (i.e. $D_1=\varepsilon=\num[round-mode=none]{0,1}$ and $D_2=1$).
	
	The closed-loop MFC controlled mass-spring-damper system can be written according to \eqref{eq:closed-loop-sys-xtildestar} and \eqref{eq:closed-loop-sys-ztilde} as
	\begin{subequations}
		\begin{align}
			\dot{\tilde{\bm{x}}}^\star &= \left(\bm{A}+\bm{b}{\bm{k}^\star}^\transpose\right) \tilde{\bm{x}}^\star \label{eq:closed-loop-mfc-msd-xtildestar} \\
			\dot{\tilde{\bm{z}}} &=  \left(\bm{A}+\bm{b}\tilde{\bm{k}}^\transpose\right) \tilde{\bm{z}} + \bm{b} \tilde{\phi}(\bm{x}_{\mathrm{d}} + \bm{D} \bm{z}_{\mathrm{s}}, \tilde{\bm{x}}^\star + \bm{D} \tilde{\bm{z}} )  \, . \label{eq:closed-loop-mfc-msd-ztilde}
		\end{align} \label{eq:closed-loop-mfc-msd}
	\end{subequations}
	
	In order to prove stability of the closed-loop MFC system, we choose the quadratic Lyapunov function according to \eqref{eq:LF-MFC} with positive definite matrix~$\bm{P}$.
	We calculate $\bm{P}$ according to \eqref{eq:LF_equation} with the control parameter \eqref{eq:control_parameter} and obtain  
	\begin{equation}\label{eq:P}
		\bm{P} = \frac{1}{32} \begin{pmatrix}
			36 & 4 \\
			4 & 5
		\end{pmatrix}.
	\end{equation}
	
	Choosing $\vartheta=100\varepsilon^{-1}=\num[round-mode=none]{1000}$ yields for the Lyapunov function in \eqref{eq:LF-MFC}
	\begin{align*}
		V(\tilde{\bm{x}}^\star, \tilde{\bm{z}}) &= \vartheta {\tilde{\bm{x}}^\star{}}^\transpose \bm{P} \tilde{\bm{x}}^\star + \tilde{\bm{z}}^\transpose \bm{P} \tilde{\bm{z}} \\
		&= \frac{1000}{32} {\tilde{\bm{x}}^\star{}}^\transpose \begin{pmatrix}
			36 & 4 \\
			4 & 5
		\end{pmatrix} \tilde{\bm{x}}^\star + \frac{1}{32} \tilde{\bm{z}}^\transpose \begin{pmatrix}
			36 & 4 \\
			4 & 5
		\end{pmatrix} \tilde{\bm{z}} \, .
	\end{align*}
	Following the analysis of Section~\ref{sec:analysis} we compute the robustness bound \eqref{eq:stabi-bedingung} for our model-following control design
	\begin{equation}\label{eq:gamma_MFC}
		\Gamma_{\mathrm{MFC}} =\frac{1}{\varepsilon \left( 1 + \sqrt{1 + (\vartheta \varepsilon)^{-1}} \right) \normtwo{\bm{b}^\transpose \bm{P}} } = \num{24,9256} \, .
	\end{equation}
	This robustness bound $\Gamma_{\mathrm{MFC}}$ represents the maximum gain for $\gamma$ in condition~\eqref{eq:Lipschitz-phitilde} as derived in \eqref{eq:stabi-bedingung}.
	Next, we use this robustness bound to calculate an estimate of the region of attraction.
	
	\subsection{Estimated region of attraction}
	\label{sec:roa-msd}
	
	In this section we determine an estimate for the region of attraction 
	for system~\eqref{eq:msd-dgl-x} with uncertainty~\eqref{eq:msd-uncertainty} and the model-following control law given by~\eqref{eq:model-controller}--\eqref{eq:process-controller} with feedback gains given in~\eqref{eq:control_parameter}.
	We shall use the stability analysis in Section \ref{sec:analysis} which is based on the estimate of the norm bound on $\tilde \phi$ in \eqref{eq:Lipschitz-phitilde}.
	Thus we need to ensure that this estimate is valid throughout the solution of the closed-loop system \eqref{eq:closed-loop-mfc-msd}.
	
	We consider two different approaches to estimate the region of attraction for the MFC scheme. 
	In the first approach, the estimated region of attraction is calculated for the combined state $({\tilde{\bm{x}}^\star{}}^\transpose, \tilde{\bm{z}}^\transpose)^\transpose$.
	Whereas in the second approach the knowledge of $\tilde{\bm{x}}^\star$ is used and the estimated region of attraction is calculated considering the constraints imposed by $\tilde\phi$ only for the state $\tilde{\bm{z}}$.

	The first approach employs the established method by considering a level-set of the Lyapunov function $V$, \cite{Khalil2002}. 
	We consider the combined state vector $({\tilde{\bm{x}}^\star{}}^\transpose, \tilde{\bm{z}}^\transpose)^\transpose$ and the estimated region of attraction is then given by the set
	\begin{equation}
		\Omega_\mathrm{MFC1}\coloneqq\left\lbrace \begin{pmatrix}
			\tilde{\bm{x}}^\star \\
			\tilde{\bm{z}}
		\end{pmatrix} \in \mathbb{R}^4  \Big| V(\tilde{\bm{x}}^\star, \tilde{\bm{z}})\leq c_1 \right\rbrace \label{eq:omega-combined-state}
	\end{equation}
	with $c_1$ to be determined. 
	As outlined above we calculate the level $c_1$ such that the estimate on the norm of $\tilde{\phi}$ satisfies \eqref{eq:Lipschitz-phitilde} while also satisfying the stability condition~\eqref{eq:stabi-bedingung}, i.e. $\gamma<\Gamma_\mathrm{MFC1}$.
	
	Consider the estimate of $\tilde{\phi}$ in \eqref{eq:msd-phi-tilde-xtildestar-ztilde}.
	Using the estimate for the scalar product of two vectors $\bm{v}_1$ and $\bm{v}_2$: $\bm{v}_1^\transpose \bm{v}_2 \leq \normtwo{\bm{v}_1} \normtwo{\bm{v}_2}$ together with some elementary calculations we obtain
	\begin{multline}\label{eq:phitilde-norm}
		\normtwo{\tilde{\phi} (\bm{x}_{\mathrm{d}}+\bm{D} \bm{z}_{\mathrm{s}}, \tilde{\bm{x}}^\star + \bm{D} \tilde{\bm{z}}) } \leq \\
		\frac{1}{m} \biggl( \Big[ \abs{\Delta k} + \bar{\sigma}_1 \Big( \left(  \normtwo{\tilde{\bm{x}}^\star} + \normtwo{\tilde{\bm{z}}} + \tfrac{3}{2} \normtwo{\bm{x}_{\mathrm{d}} + \bm{D} \bm{z}_{\mathrm{s}} } \right)^2  \\ 
		+ \tfrac{3}{4} \normtwo{\bm{x}_{\mathrm{d}} + \bm{D} \bm{z}_{\mathrm{s}} }^2 \Big) \Big]^2 + \Delta c_{\mathrm{d}}^2 \biggr)^{\frac{1}{2}}  
		\big( \normtwo{\tilde{\bm{x}}^\star} + \normtwo{\tilde{\bm{z}}} \big) ,
	\end{multline}
	where the constant $\bar\sigma_1$ is given by
	\begin{equation}
		\bar{\sigma}_1 = \abs{\Delta k}\left(\alpha + \abs{\Delta \alpha}\right)^2 + k \abs{\Delta \alpha} \left(2\alpha + \abs{\Delta \alpha}\right) \, . \label{eq:sigma1bar}
	\end{equation}
	Thus the constant $\gamma$ in \eqref{eq:Lipschitz-phitilde} 
	is given by
    \begin{align}\label{eq:msd-gamma}
				\hspace{-1em}	\gamma\coloneqq \sup_{\bm{x}\in\mathcal D}
				\frac{1}{m} \biggl( \!\Big[ \abs{\Delta k} + \bar{\sigma}_1 \Big(\!\! \left(  \normtwo{\tilde{\bm{x}}^\star} \!+\! \normtwo{\tilde{\bm{z}}} 
				+ \tfrac{3}{2} \normtwo{\bm{x}_{\mathrm{d}} \!+\! \bm{D} \bm{z}_{\mathrm{s}} } \right)^2  
				+ \tfrac{3}{4} \normtwo{\bm{x}_{\mathrm{d}} \!+\! \bm{D} \bm{z}_{\mathrm{s}} }^2 \Big) \Big]^2 + \Delta c_{\mathrm{d}}^2 \biggr)^{\frac{1}{2}},  
	\end{align}
			where the region of attraction to be determined is a subset of $\mathcal D$.
			
			Further, we 
			substitute $\normtwo{\tilde{\bm{x}}^\star} + \normtwo{\tilde{\bm{z}}}$ as a function of the Lyapunov function $V$. Therefore, we use the simplification $\frac{1}{2} \left(\normtwo{\tilde{\bm{x}}^\star} + \normtwo{\tilde{\bm{z}}}\right)^2 \leq \normtwo{\tilde{\bm{x}}^\star}^2 + \normtwo{\tilde{\bm{z}}}^2$ and the minimum estimate of the Lyapunov function (with $\vartheta \geq 1$) that satisfies
			\begin{multline*}
				\hspace{-1em}\lambda_{\mathrm{min}}(\bm{P}) \frac{1}{2} \left(\normtwo{\tilde{\bm{x}}^\star} \!+\! \normtwo{\tilde{\bm{z}}}\right)^2 \leq \lambda_{\mathrm{min}}(\bm{P}) \left(\normtwo{\tilde{\bm{x}}^\star}^2 \!+\! \normtwo{\tilde{\bm{z}}}^2\right)
				\leq \lambda_{\mathrm{min}}(\bm{P}) \left(\vartheta \normtwo{\tilde{\bm{x}}^\star}^2 \!+\! \normtwo{\tilde{\bm{z}}}^2\right) \leq V \leq c_1 
			\end{multline*}
			and solve this inequality for $\normtwo{\tilde{\bm{x}}^\star} + \normtwo{\tilde{\bm{z}}}$ to obtain
			\begin{equation}
				\normtwo{\tilde{\bm{x}}^\star} + \normtwo{\tilde{\bm{z}}} \leq \sqrt{\frac{2 c_1}{\lambda_{\mathrm{min}}(\bm{P})}} \, . \label{eq:normxstar-z-dependent-on-c}
			\end{equation}
			The maximum level $c_1$ is calculated by substituting \eqref{eq:normxstar-z-dependent-on-c} into \eqref{eq:msd-gamma}, evaluating the stability condition $\Gamma_\mathrm{MFC}>\gamma$ given by
			\begin{multline*}
				\Gamma_\mathrm{MFC} > \frac{1}{m} \Biggl( \Delta c_{\mathrm{d}}^2 + \biggl[ \abs{\Delta k}  + \bar{\sigma}_1 \biggl( \frac{3}{4} \normtwo{\bm{x}_{\mathrm{d}} + \bm{D} \bm{z}_{\mathrm{s}} }^2 
				+  \left(  \sqrt{\frac{2c_1}{\lambda_{\mathrm{min}}(\bm{P})}} + \frac{3}{2} \normtwo{\bm{x}_{\mathrm{d}} + \bm{D} \bm{z}_{\mathrm{s}} } \right)^2  
				\biggr) \biggr]^2  \Biggr)^{\frac{1}{2}}
			\end{multline*}
			and solving for $c_1$, i.e. 
			\begin{equation}\label{eq:estimateROA_approach1_c1}
				c_1 \leq \lambda_{\mathrm{min}}(\bm{P}) r_{\mathrm{MFC1}}^2
			\end{equation}  
			with
			\begin{align}
				\hspace{-1em}	r_{\mathrm{MFC1}} \!\coloneqq\! \left(\!\!\!\frac{\sqrt{\left(m\Gamma_{\mathrm{MFC}}\right)^2\!\!-\!\Delta c_{\mathrm{d}}^2 } - \abs{\Delta k} }{2 \bar{\sigma}_1 } - \frac{3}{8} \normtwo{\bm{x}_{\mathrm{d}} \!+ \!\bm{D} \bm{z}_{\mathrm{s}}}^2 \!\!\right)^{\!\!\!\frac{1}{2}}  
				- \frac{3}{2 \sqrt{2}} \normtwo{\bm{x}_{\mathrm{d}} + \bm{D} \bm{z}_{\mathrm{s}}}\, .     \label{eq:r-mfc-approach1} 
			\end{align}
			Note that a large enough $\Gamma_{\mathrm{MFC}}$ renders $r_{\mathrm{MFC1}}\geq0$ which yields a valid estimate of the region of attraction.

			The second approach to calculate an estimate of the region of attraction exploits the fact that the dynamics in the MCL are not exposed to any uncertainty or disturbance and the state  $\tilde{\bm{x}}^\star$ is perfectly known.
			By design, the closed-loop dynamics \eqref{eq:closed-loop-mfc-msd-xtildestar} for $\tilde{\bm{x}}^\star$ are globally asymptotically stable and thus independent of the validity of the condition $\Gamma>\gamma$.
			Therefore we only have to impose the constraints given by the Lipschitz condition \eqref{eq:Lipschitz-phitilde} on $\tilde{\bm{z}}$.
			
			For this matter we decompose the Lyapunov function \eqref{eq:LF-MFC} as $V(\tilde{\bm{x}}^\star,\tilde{\bm{z}}) =V^{\star}(\tilde{\bm{x}}^\star)+ \tilde{V}(\tilde{\bm{z}})$ such that 
			\begin{align*}
				V^{\star}(\tilde{\bm{x}}^\star)&:=\vartheta {\tilde{\bm{x}}^\star{}}^\transpose \bm{P} \tilde{\bm{x}}^\star \, , & \tilde{V}(\tilde{\bm{z}}) &:= \tilde{\bm{z}}^\transpose \bm{P} \tilde{\bm{z}}.
			\end{align*}%
			Consider the set
			\begin{equation} \label{eq:omega-separate-state}
				\Omega_\mathrm{MFC2}\coloneqq \left\lbrace \begin{pmatrix}
					\tilde{\bm{x}}^\star \\
					\tilde{\bm{z}}
				\end{pmatrix} \in \mathbb{R}^4  \Big| V(\tilde{\bm{x}}^\star, \tilde{\bm{z}})\leq c_2 \right\rbrace,
			\end{equation}
			where $c_2=c_2^\star+\tilde{c}_2$ with $\tilde c_2$ to be determined and
			\begin{equation}
				c_2^\star \coloneqq \vartheta {\tilde{\bm{x}}^\star_0{}}^\transpose \bm{P} \tilde{\bm{x}}^\star_0. \label{eq:cstar-t0}
			\end{equation}
			Note that $c_2^\star$ depends on the reference state $\bm{x}_{\mathrm{d}}$ and chosen initial values of the model $\bm{x}^\star_0$.
			Thus $\tilde c_2=c_2-c_2^\star$ also depends on that chosen initial state $\bm{x}^\star_0$.
			
			The estimated region of attraction \eqref{eq:omega-separate-state} for the closed-loop system~\eqref{eq:closed-loop-mfc-msd} with uncertainty~\eqref{eq:msd-uncertainty} is obtained by a level $\tilde{c}_2$ such that the estimate \eqref{eq:lipschitz-condition-uncertainty} on the norm of $\tilde{\phi}$ as well as the stability condition $\gamma<\Gamma_{\mathrm{MFC}}$ in \eqref{eq:stabi-bedingung} are satisfied on the set $\tilde{\Omega}(\bm{x}^\star_0)$ given by
			\begin{equation}\label{eq:levelset_tildeV}
				\tilde{\Omega}(\bm{x}^\star_0) \coloneqq \left\lbrace \tilde{\bm{z}} \in\mathbb{R}^2 \big| \tilde{V}(\tilde{\bm{z}}) = \tilde{\bm{z}}^\transpose \bm{P} \tilde{\bm{z}} \leq \tilde{c}_2  \right\rbrace .
			\end{equation}
			To obtain a suitable level $\tilde{c}_2$, we use the estimates	
			\begin{align*}
				\vartheta \lambda_{\mathrm{min}}(\bm{P}) \normtwo{\tilde{\bm{x}}^\star}^2 &\leq c_2^\star \, ,  &
				\lambda_{\mathrm{min}}(\bm{P}) \normtwo{\tilde{\bm{z}}}^2 &\leq \tilde{c}_2
			\end{align*}
			and obtain 
			\begin{align*}
				\normtwo{\tilde{\bm{x}}^\star} &\leq \sqrt{\frac{c_2^\star}{\vartheta \lambda_{\mathrm{min}}(\bm{P})}} \, , & 
				\normtwo{\tilde{\bm{z}}} &\leq \sqrt{\frac{\tilde{c}_2}{ \lambda_{\mathrm{min}}(\bm{P})}}.
			\end{align*}
			Substituting the estimates for the norm of the state  into~\eqref{eq:msd-gamma}, and evaluating the stability condition \mbox{$\Gamma_\mathrm{MFC}>\gamma$} 
			and solving for $\tilde{c}_2$ yields
			\begin{equation}\label{eq:estimateROA_approach2_tilde_c2}
				\tilde{c}_2 \leq \lambda_{\mathrm{min}}(\bm{P}) \tilde{r}_\mathrm{MFC2}^2
			\end{equation}
			with
			\begin{align}
				\tilde{r}_\mathrm{MFC2} \!&\coloneqq\!\! \left(\!\!\frac{\sqrt{\left(m\Gamma_{\mathrm{MFC}} \right)^2\!\!-\!\Delta c_{\mathrm{d}}^2 } - \abs{\Delta k} }{\bar{\sigma}_1 } - \frac{3}{4} \normtwo{\bm{x}_{\mathrm{d}} \!+\! \bm{D} \bm{z}_{\mathrm{s}}}^2\!\right)^{\!\!\!\!\frac{1}{2}} 
				- \frac{3}{2} \normtwo{\bm{x}_{\mathrm{d}} + \bm{D} \bm{z}_{\mathrm{s}}} - \sqrt{\frac{c_2^\star}{\vartheta \lambda_{\mathrm{min}}(\bm{P})}}  \, .    \label{eq:r-tilde-approach2}
			\end{align}
			
			\begin{remark}
				Note that we require $r_{\mathrm{MFC2}}\geq0$ for a valid estimate of the region of attraction. 
				This results in
				an upper bound for $c_2^\star$ given by
				\begin{align*}
					c_2^\star &\leq  \left(\!\left(\!\!\frac{\sqrt{\left(m\Gamma_{\mathrm{MFC}} \right)^2\!\!-\!\Delta c_{\mathrm{d}}^2 } - \abs{\Delta k} }{\bar{\sigma}_1 } - \frac{3}{4} \normtwo{\bm{x}_{\mathrm{d}} \!+\! \bm{D} \bm{z}_{\mathrm{s}}}^2\!\right)^{\!\!\!\!\frac{1}{2}} 
					- \frac{3}{2} \normtwo{\bm{x}_{\mathrm{d}} + \bm{D} \bm{z}_{\mathrm{s}}} \right) \vartheta \lambda_{\mathrm{min}}(\bm{P}) \, .
				\end{align*} 
				Large enough $\Gamma_{\mathrm{MFC}}$ renders the right-hand side positive.
			\end{remark}
			
			The set $\tilde \Omega(\bm{x}^\star_0)$ represents all initial process states $\bm{x}_0=\bm{D}\tilde{\bm{z}}_0 + \bm{x}^\star_0 + \bm{x}_{\mathrm{s}}$ for which the MFC scheme is guaranteed to converge.
			In this sense $\tilde \Omega(\bm{x}^\star_0)$ may be interpreted as uncertainty region for the true process state if its initial value is estimated to $\bm{x}^\star_0$.

		Let us compare the level-sets obtained for the two approaches to estimate the region of attraction for the MFC scheme. 
		To that effect we bring the bound for the level $c_1$ of the first approach into the form of the second approach, i.e. we consider $c_1=c_1^\star+\tilde{c}_1$. Suppose $c^\star=c_1^\star=c_2^\star$ is a fixed value for both approaches. Then for \eqref{eq:estimateROA_approach1_c1}, we obtain 
		\begin{align*}
			\tilde{c}_1 \leq \lambda_{\mathrm{min}}(\bm{P}) r_{\mathrm{MFC1}}^2 - c^\star = \lambda_{\mathrm{min}}(\bm{P}) \left(\frac{1}{\sqrt{2}} r_a\right)^2 - c^\star
		\end{align*}
		with auxiliary variable
		\begin{align*}
			\hspace{-0.8em}	
			r_a \!\coloneqq\! \left(\!\frac{\sqrt{\left(m\Gamma_{\mathrm{MFC}} \right)^2\!\!-\!\Delta c_{\mathrm{d}}^2 } - \abs{\Delta k} }{\bar{\sigma}_1 } - \frac{3}{4} \normtwo{\bm{x}_{\mathrm{d}} + \bm{D} \bm{z}_{\mathrm{s}}}^2\right)^{\!\!\!\!\frac{1}{2}} 
			- \frac{3}{2} \normtwo{\bm{x}_{\mathrm{d}} + \bm{D} \bm{z}_{\mathrm{s}}} \, .
		\end{align*}
		Whereas \eqref{eq:estimateROA_approach2_tilde_c2} gives
		\begin{align*}
			\tilde{c}_2 \leq \lambda_{\mathrm{min}}(\bm{P}) \tilde{r}_\mathrm{MFC2}^2 = \lambda_{\mathrm{min}}(\bm{P}) \left(r_a - \sqrt{\frac{c_2^\star}{\vartheta \lambda_{\mathrm{min}}(\bm{P})}}\right)^2 .
		\end{align*}
		With the above estimates and assuming $\vartheta>1$, the second approach yields the larger level since  $\tilde{c}_2 - \tilde{c}_1 > 0$ as is shown by the following calculation:
		\begin{align*}
			\lambda_{\mathrm{min}}(\bm{P}) \! \left( \! \left(r_a-\sqrt{\frac{c_2^\star}{\vartheta \lambda_{\mathrm{min}}(\bm{P})}}\right)^2 \! -\left(\frac{1}{\sqrt{2}}r_a\right)^2\right)+c^\star &> 0\\
			\left(r_a - \sqrt{\frac{c_2^\star}{\vartheta \lambda_{\mathrm{min}}(\bm{P})}}\right)^2 - \left(\frac{1}{\sqrt{2}} r_a\right)^2 + \frac{c^\star}{\lambda_{\mathrm{min}}(\bm{P})} &> 0  \\
			\left(\frac{1}{\sqrt{2}} r_a - \sqrt{\frac{2 c_2^\star}{\vartheta \lambda_{\mathrm{min}}(\bm{P})}}\right)^2 + \left(1 - \frac{1}{\vartheta}\right) \frac{c^\star}{\lambda_{\mathrm{min}}(\bm{P})} &> 0. 
		\end{align*}  
		Noting that $(1-\frac{1}{\vartheta})>0$ for $\vartheta>1$ establishes the claim.
		Thus, for the same Lyapunov function the second approach leads to a larger level-set for which $\gamma<\Gamma_{\mathrm{MFC}}$ and stability is ensured.
		
		\subsection{Robustness of the single-loop designs}
		
		For comparison we consider the single-loop designs using the model controller $\bm{k}^\star$ as well as the high-gain control~$\tilde{\bm{k}}$.
		The first single-loop control law is given by~\eqref{eq:single-loop-control-law-allg} with feedback gain $\bm{k}_{\mathrm{SL}}={\bm{k}^\star{}}$.
		The steady-state $\bm{x}_{\mathrm{s}}$ can be calculated by~\eqref{eq:single-loop-steady-state-allg}.
		We obtain $x_{\mathrm{s}, 2}=0$ and the zeros of the polynomial 
		\begin{align}
			p_{\mathrm{SL}, \mathrm{s}}(x_1)= k_1^\star (x_{1}-y_{\mathrm{d}}) - \left( \frac{\Delta k}{m} \left( \alpha + \Delta \alpha\right)^2 x_{1}^3 
			+ \frac{k}{m} \Delta a \left( 2 \alpha + \Delta \alpha \right) x_{1}^3   +\frac{\Delta k}{m} x_{1} \right) \, . \label{eq:sl-steady-state-function} 
		\end{align} 
		Note that $p_{\mathrm{SL}, \mathrm{s}}(x_{\mathrm{s}, 1})=0$ may have more than one solution. We denote by $\bm{x}_{\mathrm{s}}$ the solution that is closest to $\bm{x}_{\mathrm{d}}$.
		We calculate the error dynamics given by~\eqref{eq:single-loop-error-dynamics}, i.e. 
		\begin{equation*}
			\dot{\bm{x}}_e = (\bm{A} + \bm{b} \bm{k}_{\mathrm{SL}}^\transpose ) \bm{x}_e + \bm{b} \tilde{\phi}(\bm{x}_{\mathrm{s}}, \bm{x}_e)
		\end{equation*}
		with uncertainty $\tilde{\phi}$ given by~\eqref{eq:msd-phi-tilde-xs-xe}. Further, the quadratic Lyapunov function~\eqref{eq:single-loop-Lyapunov-function} with matrix $\bm{P}$ given by~\eqref{eq:P} and Lipschitz continuity of the model uncertainty 
		\begin{equation}
			\normtwo{\tilde{\phi}(\bm{x}_{\mathrm{s}},\bm{x}_e)} < \gamma_{\mathrm{SL}} \normtwo{\bm{x}_e}  \label{eq:phitilde-norm-sl-condition-msd-allg}
		\end{equation}
		are used for the stability analysis. 
		For stability, condition~\eqref{eq:Gamma_condition_SL} has to be satisfied with
		\begin{align}
			\Gamma_{\mathrm{SL}}=\frac{1}{2\normtwo{\bm{b}^\transpose \bm{P}}} = \num{2,4988} \, . \label{eq:Gamma-sl-msd}
		\end{align}
		Note that the robustness bound of the MFC scheme $\Gamma_{\mathrm{MFC}}$ in \eqref{eq:gamma_MFC}  is scaled by $\varepsilon^{-1}$ and thus roughly ten times larger than the robustness bound of the single-loop design $\Gamma_{\mathrm{SL}}$ which is the expected result of the analysis given by Eq.~\eqref{eq:Gamma-sl-comparison-allg}.

		
		For completeness, we calculate an estimate of the region of attraction for the single-loop design. The analysis steps are similar to the MFC region of attraction analysis, i.e. we first determine the set of the estimated region of attraction
		\begin{equation}\label{eq:levelset_SL}
			\Omega_{\mathrm{SL}} \coloneqq \left\lbrace \bm{x}_e \in\mathbb{R}^2 \Big| V_{\mathrm{SL}}(\bm{x}_e) \leq c_{\mathrm{SL}} \right\rbrace 
		\end{equation}
		such that the norm of $\tilde{\phi}$ satisfies~\eqref{eq:phitilde-norm-sl-condition-msd-allg} with stability condition $\gamma<\Gamma_{\mathrm{SL}}$ with $\Gamma_{\mathrm{SL}}$ given by~\eqref{eq:Gamma-sl-msd}.
		
		Consider the estimate of $\tilde{\phi}$ in \eqref{eq:msd-phi-tilde-xs-xe} given by
		\begin{align*}
			\normtwo{\tilde{\phi}(\bm{x}_{\mathrm{s}}, \bm{x}_e)} \leq \frac{1}{m} \bigg( \Delta c_{\mathrm{d}}^2  +  \Bigl[ \abs{\Delta k} 
			+ \bar{\sigma}_1 \left( \left( \normtwo{\bm{x}_e} + \tfrac{3}{2} \normtwo{\bm{x}_{\mathrm{s}}} \right)^2 + \tfrac{3}{4} \normtwo{\bm{x}_{\mathrm{s}}}^2 \right) \Bigr]^2  \biggr)^{\frac{1}{2}} \normtwo{\bm{x}_e}.
		\end{align*}
		The constant $\gamma_{\mathrm{SL}}$ of condition~\eqref{eq:phitilde-norm-sl-condition-msd-allg} is readily given by
		\begin{align*}
			\frac{1}{m} \sqrt{ \! \Delta c_{\mathrm{d}}^2  + \! \Bigl[ \abs{\Delta k} + \bar{\sigma}_1 \! \left( \! \left( \normtwo{\bm{x}_e} + \tfrac{3}{2} \normtwo{\bm{x}_{\mathrm{s}}} \right)^2 \!\! + \tfrac{3}{4} \normtwo{\bm{x}_{\mathrm{s}}}^2 \right) \Bigr]^{2} }. 
		\end{align*}
		The next step is to substitute $\normtwo{\bm{x}_e}$ as a function of the Lyapunov function. Therefore, we use the estimate of the Lyapunov function
		\begin{align*}
			\lambda_{\mathrm{min}}(\bm{P}) \normtwo{\bm{x}_e}^2 &\leq V_{\mathrm{SL}} \leq c_{\mathrm{SL}}  \\
			\normtwo{\bm{x}_e} &\leq \sqrt{\frac{c_{\mathrm{SL}}}{\lambda_{\mathrm{min}}(\bm{P})}}.
		\end{align*}
		Substituting in $\gamma_{\mathrm{SL}}$, evaluating $\Gamma_{\mathrm{SL}}>\gamma_{\mathrm{SL}}$, i.e. 
		\begin{align*}
			\Gamma_{\mathrm{SL}} > \frac{1}{m} \Biggl[ \Delta c_{\mathrm{d}}^2  +  \Biggl( \abs{\Delta k} 
			+ \bar{\sigma}_1 \left( \left( \sqrt{\frac{c_{\mathrm{SL}}}{\lambda_{\mathrm{min}}(\bm{P})}} + \tfrac{3}{2} \normtwo{\bm{x}_{\mathrm{s}}} \right)^2 + \tfrac{3}{4} \normtwo{\bm{x}_{\mathrm{s}}}^2 \right) \Biggr)^2  \Biggr]^{\frac{1}{2}},
		\end{align*}
		and solving for $c_{\mathrm{SL}}$ yields
		\begin{equation}\label{eq:estimateROA_SL_cSL}
			c_{\mathrm{SL}} < \lambda_{\mathrm{min}}(\bm{P}) r_{\mathrm{SL}}^2
		\end{equation}
		with 
		\begin{equation*}
			r_{\mathrm{SL}} \!\coloneqq\! \left(\!\frac{\sqrt{\left(m\Gamma_{\mathrm{SL}} \right)^2 \!\!-\! \Delta c_{\mathrm{d}}^2} - \abs{\Delta k} }{\bar{\sigma}_1 } - \frac{3}{4} \normtwo{\bm{x}_{\mathrm{s}}}^2 \!\right)^{\!\!\!\!\frac{1}{2}} \!\!- \frac{3}{2} \normtwo{\bm{x}_{\mathrm{s}}},
		\end{equation*}
		where $r_{\mathrm{SL}}\geq0$ must be satisfied for a valid estimate of the region of attraction.

		We consider now the single-loop high-gain design, i.e. control law \eqref{eq:single-loop-control-law-allg} with feedback gain $\bm{k}_{\mathrm{SL}}=\tilde{\bm{k}}$.
		Following the previous analysis of the single-loop design
		we determine the estimated region of attraction
		\begin{equation}\label{eq:levelset_SLHG}
			\Omega_{\mathrm{SLHG}} \coloneqq\left\lbrace \bm{z}_{\mathrm{SL}} \in\mathbb{R}^2 \big| V_{\mathrm{SLHG}}(\bm{z}_{\mathrm{SL}}) \leq c_{\mathrm{SLHG}} \right\rbrace
		\end{equation}
		such that the norm of $\tilde{\phi}$ satisfies 
		\begin{equation}\hspace{-1em}
			\normtwo{\tilde{\phi}(\bm{x}_{\mathrm{s}}, \bm{D}\bm{z}_{\mathrm{SL}})} < \gamma_{\mathrm{SLHG}} \normtwo{\bm{D}\bm{z}_{\mathrm{SL}}} \leq  \gamma_{\mathrm{SLHG}} \normtwo{\bm{z}_{\mathrm{SL}}} \label{eq:phitilde-norm-slhg-condition-msd-allg}
		\end{equation}
		with stability condition $\gamma_{\mathrm{SLHG}}<\Gamma_{\mathrm{SLHG}}$. The robustness bound according to \eqref{eq:Gamma-SLHG-scaled} is given by
		\begin{equation*}
			\Gamma_{\mathrm{SLHG}} = \frac{1}{2 \varepsilon \normtwo{\bm{b}^\transpose \bm{P}}} = \num{24.9878},
		\end{equation*} 
		where 
		the matrix $\bm{P}$ is given by \eqref{eq:P}.
		Similar calculations as for the standard single-loop design before yield for the maximal level for $V_\mathrm{SLHG}$
		\begin{equation*}
			c_{\mathrm{SLHG}} < \lambda_{\mathrm{min}}(\bm{P}) r_{\mathrm{SLHG}}^2
		\end{equation*}
		with
		\begin{align*}
			r_{\mathrm{SLHG}} \coloneqq\! \left(\frac{\sqrt{\left(m\Gamma_{\mathrm{SLHG}} \right)^2 \!-\! \Delta c_{\mathrm{d}}^2} - \abs{\Delta k} }{\bar{\sigma}_1 } - \frac{3}{4} \normtwo{\bm{x}_{\mathrm{s}}}^2 \right)^{\!\!\!\!\frac{1}{2}} 
			 - \frac{3}{2} \normtwo{\bm{x}_{\mathrm{s}}}\, .
		\end{align*}
		Note that a large enough $\Gamma_{\mathrm{SLHG}}$ renders $r_{\mathrm{SLHG}}>0$ yielding a valid estimate of the region of attraction.

		We observe that the maximum levels for the single-loop design $c_{\mathrm{SL}}$ and for the single-loop high-gain design $c_{\mathrm{SLHG}}$  only differ in the auxiliary variable $r_{\mathrm{SL}}$ and $r_{\mathrm{SLHG}}$, where the different robustness bounds $\Gamma_{\mathrm{SL}}$  and $\Gamma_{\mathrm{SLHG}}$ enter, respectively. 
		Note however that the maximum level for the single-loop design $c_{\mathrm{SLHG}}$ is related to the scaled state $\bm{z}_{\mathrm{SL}}$ and thus not directly comparable to  $c_{\mathrm{SL}}$.

		\subsection{Comparison of the designs}
		
		A comparison of these estimates of the region of attraction is not straight forward as they depend on various different parameters.
		The estimates \eqref{eq:estimateROA_approach1_c1} and \eqref{eq:estimateROA_approach2_tilde_c2}  depend on the reference state $\bm{x}_{\mathrm{d}}$. That means for the MFC scheme the estimate depends on the states $\bm{x}_{\mathrm{d}}$, $\bm{z}_{\mathrm{s}}$, and for the second approach additionally on $\tilde{\bm{x}}^\star$. For the single-loop design the estimate \eqref{eq:estimateROA_SL_cSL} depends on the steady state $\bm{x}_{\mathrm{s}}$. 
		
		We used two different approaches to estimate the region of attraction for the MFC scheme. The first approach requires 
		the estimate of the norm bound on $\tilde \phi$ in \eqref{eq:Lipschitz-phitilde} to hold for the combined state vector, whereas the second approach estimates the region of attraction applying this requirement only for the state $\tilde{\bm{z}}$. 
		This is possible since the closed MCL is linear, perfectly known without uncertainties and disturbances, and the use of the quadratic Lyapunov function $V^\star=\vartheta {\tilde{\bm{x}}^\star{}}^\transpose \bm{P} \tilde{\bm{x}}^\star$. Hence, the MCL dynamics are globally exponentially stable. 
		The level-set $V^\star(\tilde{\bm{x}}^\star{})=\vartheta {\tilde{\bm{x}}^\star{}_0}^\transpose \bm{P} \tilde{\bm{x}}^\star_0=c^\star_2$ is chosen depending on the initial state $\tilde{\bm{x}}^\star_0$, see \eqref{eq:cstar-t0}.
		Since the solution $\tilde{\bm{x}}^\star{}$ is exponentially decreasing the Lyapunov function $V^\star(\tilde{\bm{x}}^\star{})$ has its maximum value $c^\star_2$ at $t=0$.
		The set of process states $\tilde \Omega(\bm{x}^\star_0)$ given by the initial level set $\tilde{V}(\tilde{\bm{z}}_0)=\tilde{c}_2$ depends on $\tilde{\bm{x}}^\star_0$ and moves along the solution $\tilde{\bm{x}}^\star(t)$ of the MCL for $t\geq 0$, see  \cite{TietzeWR2024Anif} for an illustration.
		In fact, the value of $\tilde c_2$ may be increasing over time for a given fixed $c_2=V(\tilde{\bm{x}}^\star_0, \tilde{\bm{z}}_0)=c^\star_2+\tilde c_2$, since $c^\star_2$ decreases exponentially to zero for $t\to\infty$.
		This shall be subject of future investigations.
		
		The examples in the next section illustrate the properties of each estimate. 
		In most of the analysed cases it turns out that the MFC scheme shows an advantage in control performance and also the estimated region of attraction compared to the single-loop design. 
		
		Note that the price to be paid for the enlarged estimated region of attraction in the MFC scheme is the larger control effort when the scaled error state $\tilde{\bm{z}}$ is large.
		However, the initialisation $\bm{x}^\star_0$ of the MCL yields an additional degree of freedom compared to a single-loop high-gain design.
		Choosing $\bm{x}^\star_0$ close to $\bm{x}_{\mathrm{d}}$ approximates the single-loop high-gain design.
		Whereas the initialisation of $\bm{x}^\star_0$ close to the process $\bm{x}_0$ such that $\tilde{\bm{x}}_0$ is small reduces the initial control effort significantly such that the so-called peaking phenomenon can be avoided \cite{TietzeWR2024}.

\section{Numerical results}
\label{sec:numerical-results}

In this section we discuss simulation results for two scenarios and  the influence of the uncertainty on the steady-state error. The system parameters and uncertainties are given in Table~\ref{tab:systemparameter}.
	
	\begin{table}[tb]
		\centering
		\sisetup{per-mode=fraction,round-mode=none}
		\begin{tabular}{l*5{c}}
			\toprule
			Parameter  & $k$ & $c_{\mathrm{d}}$ & $\alpha$ & $m$ & $g_0$ \\
			Value       & \num{1.5} & \num{0,3} & \num{0,5} & \num{1} & \num{9.81} \\
			Unit    & \unit{\newton\per\meter} & \unit{\kilogram\per\second} & \unit{\per\meter\squared} & \unit{\kilogram} & \unit{\meter\per\second\squared} \\
			\cmidrule(r){1-1} \cmidrule(l){2-6}
			Uncertainty   & $\Delta k$ & $\Delta c_{\mathrm{d}}$ & $\Delta \alpha$ &  &  \\
			Value  & \num{-0.075} & \num{0,06} & \num{-0,1} & &   \\
			Unit & \unit{\newton\per\meter} & \unit{\kilogram\per\second} & \unit{\per\meter\squared} & &   \\
			\bottomrule
		\end{tabular}
		\caption{System parameters.}
		\label{tab:systemparameter}
	\end{table}
	
	The control design follows the description in Section~\ref{sec:reglerentwurf-mfc} for the considered mass-spring-damper system~\eqref{eq:msd-dgl-x} with uncertainty~\eqref{eq:msd-uncertainty}. 
	The desired eigenvalues of the MCL dynamics are placed at $\lambda_{1,2}=-2$ resulting in the state feedback vectors $\bm{\bm{k}}^\star$ and $\tilde{\bm{k}}$ given by~\eqref{eq:control_parameter}, the weight $\vartheta=100\varepsilon^{-1}$ and the time-scaling parameter $\varepsilon$ is chosen as $\varepsilon=\num[round-mode=none]{0.1}$.
	The MFC closed-loop dynamics according to~\eqref{eq:closed-loop-mfc-msd} can be calculated for specific desired outputs $y_{\mathrm{d}}$ as discussed in Section~\ref{sec:case-study}.

	\begin{figure*}[tb]
		\centering
		\includegraphics[width=\columnwidth]{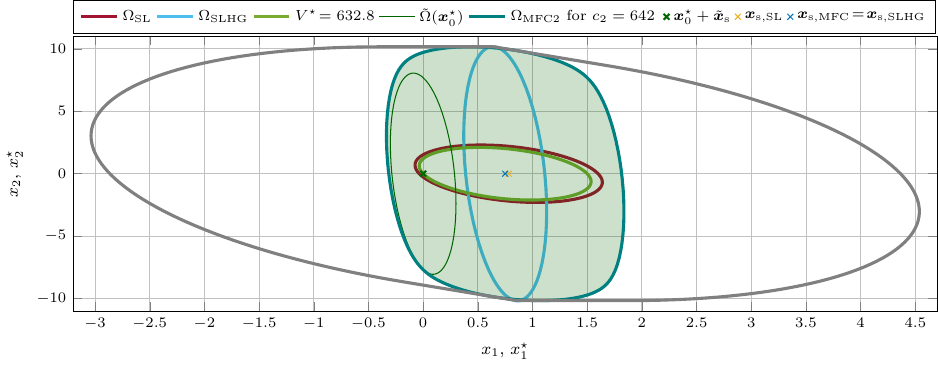}
		\caption{Estimated regions of attraction for $\bm{x}_{\mathrm{d}}=(\num[round-mode=none]{0.75} \; 0)^\transpose$ and $\bm{x}_0=\bm{0}$ and various control designs.}
		\label{fig:phasenportrait-xd-075-ROAs}
	\end{figure*}

\begin{figure}[t]
		\centering
		\includegraphics{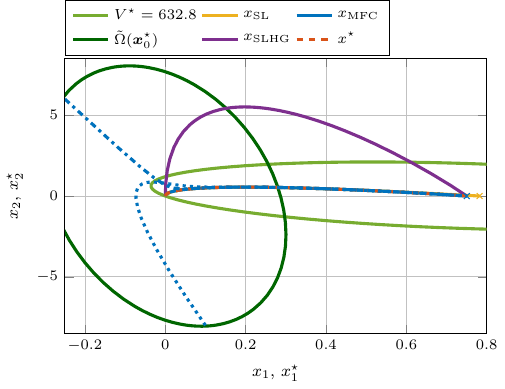}
		\caption{Phase portrait and estimated region of attraction for $\bm{x}_{\mathrm{d}}=(\num[round-mode=none]{0.75} \; 0)^\transpose$ and $\bm{x}_0=\bm{0}$.  For the MFC scheme we have $\bm{x}^\star_0=\bm{0}$ with three different initial values of the process state: $\bm{x}_0=\bm{0}$ (solid blue line), $\bm{x}_0=(0.1 \enspace -8)^\transpose$ (dotted blue line), and $\bm{x}_0=(-0.25 \enspace 6)^\transpose$ (dashed-dotted blue line).}
		\label{fig:phasenportrait-xd-075}
	\end{figure}

\begin{figure}[h]
		\centering 
		\includegraphics{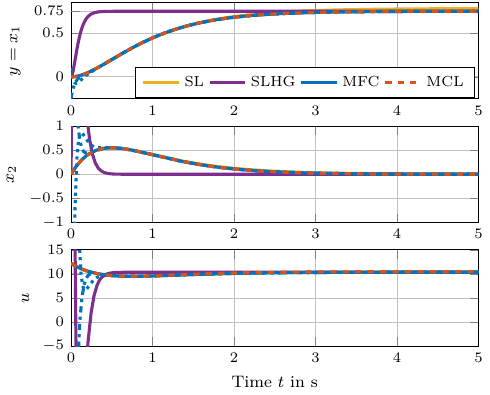}		
		\vspace{-1.5ex}
		\caption{Time-responses of the states and input signal for $\bm{x}_{\mathrm{d}}=(\num[round-mode=none]{0.75} \; 0)^\transpose$ and  $\bm{x}_0=\bm{0}$. For the MFC scheme we have $\bm{x}^\star_0=\bm{0}$ with three different initial values of the process states: $\bm{x}_0=\bm{0}$ (solid blue line), $\bm{x}_0=(0.1 \enspace -8)^\transpose$ (dotted blue line), and $\bm{x}_0=(-0.25 \enspace 6)^\transpose$ (dashed-dotted blue line).}
		\label{fig:timeseries-xd-075}
	\end{figure}

	The first simulation scenario considers the reference set-point $\bm{x}_{\mathrm{d}}=(\num[round-mode=none]{0.75} \enspace 0)^\transpose$ with the initial state at the origin, i.e. $\bm{x}_0=\bm{x}^\star_0=\bm{0}$.
	Figures~\ref{fig:phasenportrait-xd-075-ROAs} and~\ref{fig:phasenportrait-xd-075} show the estimated regions of attraction and the phaseportraits (both in $x_1$ and $x_2$ coordinates). The time responses are depicted in Figure~\ref{fig:timeseries-xd-075}.
	
	In Figure~\ref{fig:phasenportrait-xd-075-ROAs} the ruby ellipse is the estimated region of attraction $\Omega_\mathrm{SL}$ for the single-loop design in \eqref{eq:levelset_SL} obtained from the maximum level-set $V_{\mathrm{SL}}=c_{\mathrm{SL}}=0.75$ satisfying \eqref{eq:estimateROA_SL_cSL}.
	The ellipse is centred at the steady-state~$\bm{x}_\mathrm{s}$ (marked in yellow) which has an offset to the set-point~$\bm{x}_\mathrm{d}$ (not included but very close to $\bm{x}_\mathrm{s,MFC}$ marked in blue) due to the uncertainty. 
	The lightblue ellipse given by $V_{\mathrm{SLHG}}=c_{\mathrm{SLHG}}=\num[round-precision=1]{14.74}$ depicts the estimated region of attraction $\Omega_{\mathrm{SLHG}}$ for the single-loop high-gain design \eqref{eq:levelset_SLHG}. 
	Note that the initial condition $\bm{x}_0=\bm{0}$ does not lie within the estimated region of attraction of the single-loop high-gain design, thus we have no guarantee of convergence for this scenario.
	The green ellipse centred at~$\bm{x}_\mathrm{d}$ shows the level-set of the Lyapunov function $V^\star$ for the MCL with $c^\star= \num[round-precision=1]{632.813}$.
	The level-set is chosen according to \eqref{eq:cstar-t0} such that it just covers the initial model state $\bm{x}^\star_0=\bm{0}$, i.e. $c^\star=\vartheta \tilde{\bm{x}}^{\star\transpose}_0 \bm{P} \tilde{\bm{x}}^\star_0$.
	
	In dark green we have the maximum set $\tilde \Omega(\bm{x}^\star_0)$ in \eqref{eq:levelset_tildeV} satisfying \eqref{eq:estimateROA_approach2_tilde_c2} for this choice of $c^\star$ and $\bm{x}^\star_0=\bm{0}$ given by $\tilde V=\tilde c=9.2$.
	Note that $\tilde \Omega(\bm{x}^\star_0)$ is defined for $\tilde{\bm{z}}$, which describes an ellipse in the $(x_1,x_2)$-plane centred at 
	$\bm{x}_0^\star + \tilde{\bm{x}}_{\mathrm{s}}$ for the initial state.
	Thus for any initial process state $\bm{x}_0\in\tilde \Omega(\bm{x}^\star_0)$ 
	we can guarantee asymptotic stability if we choose $\bm{x}^\star_0=\bm{0}$. 
	In this sense the set $\tilde \Omega(\bm{x}^\star_0)$ can be interpreted as robustness of the MFC with respect to an uncertain initial process state~$\bm{x}_0$ for the chosen initial model state~$\bm{x}^\star_0$.
	While $t>0$ increases, the centre of~$\tilde \Omega$  moves along the solution $\bm{x}^\star(t)$ of the MCL offset by stationary error $\tilde{\bm{x}}_{\mathrm{s}}$, and represents the robustness of the MFC scheme w.r.t. the perturbed process state at any time instant $t\geq 0$ and also a bound for $\bm{x}(t)$, $t\geq 0$ in the closed-loop MFC scheme.
	
	Of course different choices of $\bm{x}^\star_0$ yield different regions~$\tilde \Omega(\bm{x}^\star_0)$ for the initial process state $\bm{x}_0$.
	Considering $\bm{x}^\star_0$ as design parameter of the MFC restricted to $V^\star(\bm x^\star_0)= \num[round-precision=1]{632.813}$ we can move the centre of $\tilde \Omega(\bm{x}^\star_0)$ around the ellipse $V^\star$ offset by~$\tilde{\bm{x}}_{\mathrm{s}}$,
	yielding the green-shaded area $\Omega_\mathrm{MFC2}$ for
	$c^\star+\tilde{c}=\num[round-precision=1]{642,045}$. 
	This represents the region of attraction for the MFC scheme for $V^\star(\bm x^\star_0)= \num[round-precision=1]{632.813}$,
	resulting in a much larger region than both single-loop designs.
	
	We may also vary the choice of $c^\star$, i.e. the level of $V^\star(\bm x^\star_0)$.
	We note that $c^\star$ decreases as $\bm{x}^\star_0$ approaches the set-point $\bm{x}_\mathrm{d}$, which allows for an increase of $\tilde c$ and thus the robustness w.r.t. the initial process state $\tilde \Omega(\bm{x}^\star_0)$ increases. 
	For $c^\star=0$, i.e. $\bm{x}^\star_0=\bm{x}_{\mathrm{d}}$, and with the weight $\vartheta\rightarrow\infty$ we get $V_{\mathrm{MFC}}=\tilde{V}=V_\mathrm{SLHG}$  as a special case. 
	If we consider all possible pairs $c^\star, \tilde c$, and thus initialisations $\bm{x}^\star_0$ for the MCL,  satisfying \eqref{eq:omega-separate-state}, \eqref{eq:cstar-t0} and\eqref{eq:estimateROA_approach2_tilde_c2}, we obtain the region of attraction for the initial process state $\bm{x}_0$ depicted by the grey line.
	This region is again considerably larger than both single-loop designs.
	Note, however, that some of these initial state $\bm{x}_0$ require perfect knowledge of the initial process state such that $\bm{x}^\star_0=\bm{x}_0$ can be chosen.
	But in many cases we have a significant robustness with respect to the uncertain initial state $\bm{x}_0$ yielding again a much larger region of attraction.

	Figure~\ref{fig:phasenportrait-xd-075} shows several trajectories for $\bm{x}_0=\bm{0}$ and two additional trajectories for $\bm{x}_0\neq \bm{0}$.
	The yellow line (mostly covered) and the purple line show the single-loop simulations.
	Remarkably, the single-loop high-gain approach shows convergence even though we have not shown stability for this initial condition.
	The yellow line is covered by the MFC solution $\bm{x}^\star$ (dashed red) as the MCL uses exactly the same control law as the single-loop design.
	Solid blue is the process state $\bm{x}$ for the MFC high-gain design which follows $\bm{x}^\star$ very closely.
	Additionally, we have included two further simulations of the MFC with perturbed initial process state in dashed and dash-dotted blue.
	This illustrates the robustness of the MFC towards uncertain initial states.
	
	The time response in Figure \ref{fig:timeseries-xd-075}
	shows that the MFC solutions are very similar to the standard single-loop design. 
	However, the single-loop design has a large steady-state error of about \qty[round-mode=none]{4,3}{\percent} whereas the steady-state error of the MFC scheme is negligible. 
	We also observe that the deviations from the initial process state (dashed and dash-dotted lines) are rapidly compensated in the MFC scheme (less than \qty[round-mode=none]{0.5}{\second}).
	The single-loop high-gain design (purple) shows a much faster convergence with identical (negligible  steady-state error) at the expense of very large control effort $u_{\mathrm{SLHG}}(0)\approx \num[round-mode=none]{310}$.
	The MFC with the same initial process state only requires $u_{\mathrm{MFC}}(0)\approx \num[round-mode=none]{13}$.
	In this sense the MFC inherits the benefits of both single-loop designs: the steady-state performance of the high-gain design with moderate control effort of the simple single-loop design.
	Only if we move the initial process state to the extreme boundary of the estimated region of attraction we observe large control effort of $u_{\mathrm{MFC}}(0) \approx -127$ for $\bm{x}_0=(-0.25 \enspace 6)^\transpose$, and even $u_{\mathrm{MFC}}(0)\approx 290$ for  $\bm{x}_0=(0.1 \enspace -8)^\transpose$.

	\begin{figure}[h!]
		\centering 
		\includegraphics{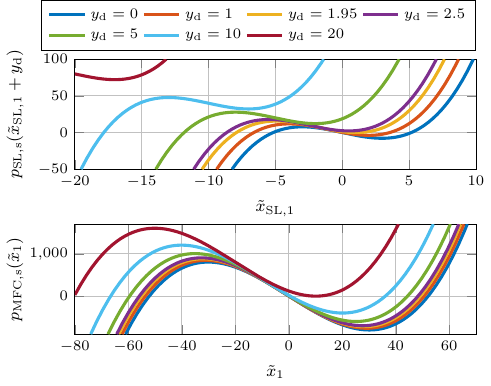}
		\vspace{-3ex}
		\caption{Functions $p_{\mathrm{SL}, \mathrm{s}}(x_1)$ and $p_{\mathrm{MFC}, \mathrm{s}}(\tilde{x}_1)$. They show the steady-state for $p_{\mathrm{SL}, \mathrm{s}}(x_{\mathrm{s}, 1})=0$ and $p_{\mathrm{MFC}, \mathrm{s}}(\tilde{x}_{\mathrm{s}, 1})=0$.}
		\label{fig:steady-state-solution}
	\end{figure}
	\begin{figure}[t]
		\centering 
		\includegraphics{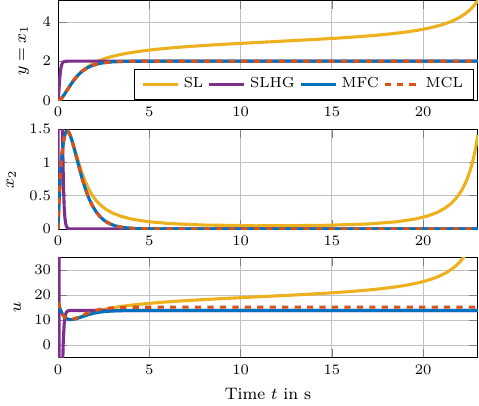}
		\vspace{-.5ex}
		\caption{Time-response of states and input signal for $\bm{x}_{\mathrm{d}}=(\num[round-mode=none]{2} \; 0)^\transpose$ and  $\bm{x}_0=\bm{x}^\star_0=\bm{0}$.}
		\label{fig:timeseries-xd-2}
	\end{figure}
	
	\begin{figure}[t]
		\centering
		\includegraphics{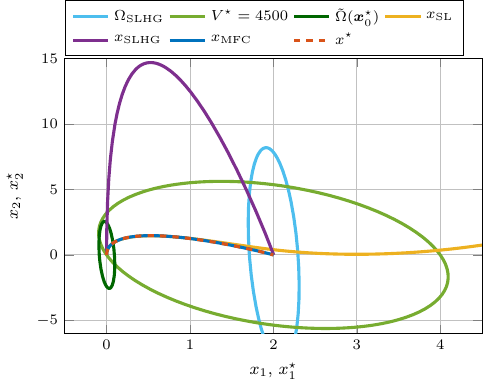}
		\vspace{-2ex}
		\caption{Phase portrait and estimated region of attraction for $\bm{x}_{\mathrm{d}}=(\num[round-mode=none]{2} \; 0)^\transpose$ and  $\bm{x}_0=\bm{x}^\star_0=\bm{0}$.}
		\label{fig:phasenportrait-xd-2}
	\end{figure}
	
	The second scenario considers the reference state at $\bm{x}_{\mathrm{d}}=(\num[round-mode=none]{2} \enspace 0)^\transpose$ and the initial state at $\bm{x}_0=\bm{x}^\star_0=\bm{0}$. 
	For the single-loop design the uncertainty disturbs the steady-state \eqref{eq:single-loop-steady-state-allg}  such that the set-point ceases to be an equilibrium.
	Defining $\tilde{\bm{x}}_{\mathrm{SL}}\coloneqq \bm{x} - \bm{x}_{\mathrm{d}}$ and substituting $x_1$ in Eq.~\eqref{eq:sl-steady-state-function} we obtain
	\begin{multline*}\hspace{-0.5em}
		p_{\mathrm{SL}}(\tilde{x}_{\mathrm{SL}, 1} + y_{\mathrm{d}}) \!=\! k_1^\star \tilde{x}_{\mathrm{SL}, 1} - \left(\! \frac{\Delta k}{m} \left( \alpha + \Delta \alpha\right)^2 (\tilde{x}_{\mathrm{SL}, 1}+y_{\mathrm{d}})^3 \right. \\
		\left. + \frac{k}{m} \Delta a \left( 2 \alpha + \Delta \alpha \right) (\tilde{x}_{\mathrm{SL}, 1} + y_{\mathrm{d}})^3   +\frac{\Delta k}{m} (\tilde{x}_{\mathrm{SL}, 1} + y_{\mathrm{d}}) \right).
	\end{multline*}
	The top plot of Figure \ref{fig:steady-state-solution} shows this polynomial for various values of the desired output $y_\mathrm{d}$.
	We observe that there are up to three possible solutions for the steady-state equation.
	However for a desired output of $y_{\mathrm{d}}>\num[round-mode=none]{1,95}$, there remains only one solution for the considered uncertainty. 
	The remaining steady-state is unstable.
	The plot at the bottom of Figure~\ref{fig:steady-state-solution} depicts the function $p_{\mathrm{MFC}}(\tilde{x}_1)$ given by Eq.~\eqref{eq:mfc-steady-state-function} for the MFC scheme. 
	We observe that the stable equilibria in terms of the error state $\tilde x_1$ remain close to zero for significantly larger values of $y_\mathrm{d}$. They are accompanied by unstable equilibria that are much further away than in the single-loop design.
	Note that the steady-state results of the MFC design are exactly the same for the single-loop high-gain design as discussed in Section~\ref{sec:analyse-sl-slhg}.

	Accordingly the single-loop design is not able to stabilise the reference state whereas the single-loop high-gain and the MFC high-gain design reach steady-state with negligible error, see Figure \ref{fig:timeseries-xd-2}.
	Again the MFC requires significantly less control effort than the single-loop high-gain control $u(0)\approx \num[round-mode=none]{810}$.
	Figure~\ref{fig:phasenportrait-xd-2} shows the phaseplane with the estimated regions of attraction.
	The initial condition $\bm{x}_0=\bm{0}$ is not within the estimated the region of attraction for the single-loop high-gain design (lightblue).
	The robustness with respect to deviations of the initial conditions $\bm{x}_0$ and $\bm{x}^\star_0$ given by $\tilde V$ is smaller than in the first scenario.
	This is caused by a larger deviation of the model state to the reference state which requires a larger level-set $V^\star$ and allows only for smaller values of $\tilde c$ for the level-set of $\tilde V$.

\section{Conclusions}
\label{sec:conclusions}

We study the MFC architecture for nonlinear systems with flat output subject to locally Lipschitz uncertainties.
We provide a thorough analysis of the region of attraction provided by a quadratic Lyapunov function approach.
The case study demonstrates the control design and shows that a much larger region of attraction can be guaranteed by the MFC scheme compared to the same design in a single-loop.
In fact the MFC scheme achieves the same steady-state precision as the single-loop
high-gain
design, but with less control effort and larger region of attraction.
The model-following control scheme combines the moderate transient response and control effort of the standard single-loop design with the increased robustness of the high-gain approach.
Additionally we can guarantee a much larger region of attraction than for the single-loop high-gain design.
Note that the MFC approach can be easily adapted to nonlinear systems with internal dynamics and no flat output.

\bibliographystyle{plain}
\bibliography{referenzen}

\end{document}